\documentclass[a4paper,12pt]{article}
\pdfoutput=1
\usepackage{amsmath,amssymb,amsthm,amsxtra,overpic,bbm,bbold,bm,epsfig}
\usepackage{color,ulem,subfigure,hyperref}

\usepackage[latin1]{inputenc}
\usepackage{ae,aecompl}
\usepackage{amsmath,amssymb,mathrsfs}
\usepackage{graphicx}
\usepackage{multirow} 
\usepackage{rotating}

\usepackage{graphicx}
\usepackage{mathtools}
\usepackage{amssymb}
\usepackage{amsfonts}
\usepackage[footnotesize]{caption}
\usepackage{color}
\usepackage{braket}
\usepackage{cite}
\usepackage{hyperref}
\usepackage{url}
\usepackage{multirow}
\usepackage{relsize}
\usepackage{fullpage}
\usepackage{makecell}
\usepackage{blkarray}
\usepackage{fullpage}
\usepackage{bbm,bbold}

\usepackage{url}

\allowdisplaybreaks

\newcommand{\Lmutau}{\Lambda_{\mu\tau}}
\newcommand{\Lew}{\Lambda_{\text{EW}}}
\newcommand{\Lss}{\Lambda_{\text{ss}}}

\allowdisplaybreaks

\begin{document}

\begin{titlepage}

\begin{center}
{\bf\Large
\boldmath{
Littlest mu-tau seesaw
}
} 
\\[12mm]
Stephen~F.~King%
\footnote{E-mail: \texttt{king@soton.ac.uk}},
Ye-Ling~Zhou
\footnote{E-mail: \texttt{ye-ling.zhou@soton.ac.uk}},
\\[-2mm]
\end{center}
\vspace*{0.50cm}
\centerline{\it
Physics and Astronomy, University of Southampton,}
\centerline{\it
Southampton SO17 1BJ, United Kingdom }

\vspace*{1.20cm}

\begin{abstract}
{\noindent
We propose a $\mu-\tau$ reflection symmetric Littlest Seesaw ($\mu\tau$-LSS) model.
In this model the two mass parameters of the LSS model are fixed to be in a special 
ratio by symmetry, so that the resulting neutrino mass matrix in the flavour basis
(after the seesaw mechanism has been applied) satisfies
$\mu-\tau$ reflection symmetry and 
has only one free adjustable parameter, namely an overall free mass scale. 
However the physical low energy predictions of the neutrino masses and lepton mixing angles and CP phases are subject to renormalisation group (RG) corrections, which introduces further parameters.
Although the high energy model is rather complicated, involving $(S_4\times U(1))^2$ and supersymmetry,
with many flavons and driving fields, the low energy neutrino mass matrix 
has ultimate simplicity.
}
\end{abstract}
\end{titlepage}

\section{Introduction}
The nature of neutrino mass and lepton flavour mixing remains an intriguing puzzle~\cite{Xing:2011zza,King:2013eh, King:2015aea}, even as the parameters are being measured to increasing precision~\cite{Capozzi:2018ubv,deSalas:2017kay,Esteban:2016qun}. Indeed 
the latest neutrino data is consistent with the hypothesis of maximal atmospheric mixing and maximal CP violation in the lepton sector, corresponding to 
a  $\mu-\tau$ reflection symmetry, namely $\nu_{\mu} \leftrightarrow \nu_{\tau}^*$,
where the star indicates CP conjugation.
For a review of  $\mu\tau$ symmetry see e.g.\cite{Xing:2015fdg} and references therein.

The smallness of neutrino mass could originate from 
the seesaw mechanism~\cite{Minkowski:1977sc, Yanagida:1979as, GellMann:1980vs, Glashow:1979nm, Mohapatra:1979ia}.
The minimal version, involving just 
two right-handed neutrinos (2RHN), was first proposed by one of us~\cite{King:1999mb, King:2002nf}.
In such a scheme the lightest neutrino is massless. 
Such a model with two texture zeros in the Dirac neutrino mass matrix~\cite{Frampton:2002qc}
is consistent with cosmological leptogenesis~\cite{Fukugita:1986hr,Guo:2003cc, Ibarra:2003up, Mei:2003gn, Guo:2006qa, Antusch:2011nz,
Harigaya:2012bw, Zhang:2015tea}, however it is incompatible with the 
normal hierarchy (NH) of neutrino masses, 
favoured by current data~\cite{Harigaya:2012bw, Zhang:2015tea}.
On the other hand the 2RHN model with the more generic
one texture zero, as originally proposed in~\cite{King:1999mb, King:2002nf}, 
is still compatible with data in the NH.

The Littlest Seesaw (LSS) model is an example of a   
2RHN model with one texture zero, involving 
just two independent
Yukawa couplings~\cite{King:2013iva, Bjorkeroth:2014vha, King:2015dvf,Bjorkeroth:2015ora, Bjorkeroth:2015tsa,King:2016yvg,Ballett:2016yod},
leading to a highly predictive scheme characterised by 
near maximal atmospheric mixing and CP violation, as in $\mu-\tau$ reflection symmetry,
but with additional predictions arising from tri-maximal nature of the first column of the PMNS matrix
as well as a predicted reactor angle. 

In a recent paper, the LSS model was shown to respect an approximate $\mu-\tau$ reflection symmetry,
near the best fit region of parameter space, which was responsible for its approximate predictions of 
maximal atmospheric mixing and maximal CP violation in the lepton sector \cite{King:2018kka}.
This was due to a ratio of input mass parameters of the LSS having a value close to that in which the model satisfied exact $\mu-\tau$ reflection symmetry, however no model which explained this apparent coincidence was proposed.

In the present paper we shall propose a version of the LSS model which satisfies exact 
$\mu-\tau$ reflection symmetry, which we refer to as the $\mu\tau$-LSS model. We construct a supersymmetric model in the flavour symmetry $S_{4L} \times S_{4R} \times U(1) \times U(1)'$. We implement the idea of bi-multiplet to the non-Abelian discrete symmetries $S_{4L} \times S_{4R}$. Multiplets in $S_{4L}$ are used to determine the direction in the flavour space, and multiplets in $S_{4R}$ are crucial to fix the ratio of two right-handed neutrino masses and further fix the desired ratio of the two mass parameters of the LSS model. Two $U(1)$'s are imposed to give hierarchical and diagonal charged lepton masses and also to forbid unnecessary terms in the superpotential. 
The resulting neutrino mass matrix (after the seesaw mechanism has been applied) 
satisfies $\mu-\tau$ reflection symmetry and 
has only one free adjustable parameter, namely an overall free mass scale. 
However the physical low energy predictions of the neutrino masses and lepton mixing angles and CP phases are subject to renormalisation group (RG) corrections, which introduces further parameters.
Although the high energy model is rather complicated, involving $(S_4\times U(1))^2$ and supersymmetry (SUSY),
with many flavons and driving fields, it leads to a neutrino mass matrix of ultimate simplicity and beauty capable of explaining all neutrino data in terms of one adjustable overall mass scale.

The layout of the remainder of the paper is as follows. In section~\ref{sec:mutau_LSS}, we briefly review the $\mu\tau$-LSS mass texture and its prediction of oscillation parameters. In section~\ref{sec:RG}, we consider corrections of these parameters by including radiative corrections for the first time. The concrete model is given in section~\ref{sec:model}, where all flavon vacuum alignments are realised explicitly. Section~\ref{sec:conclusion} is devoted to conclusions. In the appendices, we list the basis of $S_4$ used for model building and discuss the vacuum degeneracy of flavons.

\section{The $\mu\tau$-LSS mass matrix \label{sec:mutau_LSS}} 

There are two cases of the LSS neutrino mass matrix \cite{King:2016yvg}
(after the seesaw mechanism has been implemented) namely,
\begin{eqnarray}\label{eq:mutau_mass}
\text{Case I: } & M_\nu=
\omega m_a
\begin{pmatrix}
0 & 0 & 0 \\
0 & 1 & 1 \\
0 & 1 & 1
\end{pmatrix}
+m_s
\begin{pmatrix}
1 & 3 & 1\\
3 & 9 & 3\\
1 & 3 & 1
\end{pmatrix}
 \,, \nonumber\\
\text{Case II: } & M_\nu=
\omega^2 m_a
\begin{pmatrix}
0 & 0 & 0 \\
0 & 1 & 1 \\
0 & 1 & 1
\end{pmatrix}
+m_s
\begin{pmatrix}
1 & 1 & 3\\
1 & 1 & 3\\
3 & 3 & 9
\end{pmatrix}
 \,.
\end{eqnarray} 
where $\omega = e^{i 2\pi /3}$.
As observed in \cite{King:2018kka}, if $m_{a,s}$ 
satisfy the special ratio $\frac{m_a}{m_s}=11$ then this results in maximal atmospheric mixing and CP violation,
as can be checked explicitly using the 
analytic formulas in Refs.\,\cite{King:2015dvf,King:2016yvg}.
Inserting this ratio of masses, the neutrino mass matrix takes one of the two forms
\begin{eqnarray}\label{eq:mutau_mass}
\text{Case I: } & M_\nu=m_s \left(
\begin{array}{ccc}
 1 & 3 & 1 \\
 3 & 9+11 \omega  & 3+11 \omega  \\
 1 & 3+11 \omega  & 1+11 \omega  \\
\end{array}
\right) \,, \nonumber\\
\text{Case II: } & M_\nu=m_s \left(
\begin{array}{ccc}
 1 & 1 & 3 \\
 1 & 1+11 \omega ^2 & 3+11 \omega ^2 \\
 3 & 3+11 \omega ^2 & 9+11 \omega ^2 \\
\end{array}
\right) \,.
\end{eqnarray} 
We refer to them as the $\mu\tau$-LSS mass matrices.
With the $\mu\tau$ conjugation \cite{King:2018kka}, 
\begin{eqnarray}
\nu_e \to \nu_e^* \,, \quad
\nu_\mu \to \nu_\tau^* \,, \quad
\nu_\tau \to \nu_\mu^* \,, 
\end{eqnarray}
one transforms the mass matrix from one case to the other. 
Both cases predict the same mixing angles ($\theta_{13}$, $\theta_{12}$, $\theta_{23}$), the same Dirac-type CP-violating phase ($\delta$)
\begin{eqnarray} \label{eq:mixing}
\theta_{13} &=& \arcsin \left(\frac{c_-}{\sqrt{6}}\right) \approx 7.807^\circ \,,\nonumber\\
\theta_{12} &=& \arctan \left( \frac{c_+}{2} \right) \approx 34.50^\circ \,,\nonumber\\
\theta_{23} &=& 45^\circ \,,\nonumber\\
\delta &=& 270^\circ \,,
\end{eqnarray}
where $c_- = \sqrt{1-\frac{11}{3 \sqrt{17}}}$ and $c_+ = \sqrt{1+\frac{11}{3 \sqrt{17}}}$,
and the same neutrino masses
\begin{eqnarray} \label{eq:mass}
m_1 &=& 0 \,,\nonumber\\
m_2 &=& \sqrt{\Delta m^2_{21}} = \sqrt{\frac{33}{2} \left(13-3 \sqrt{17}\right)} m_s \approx 3.226\, m_s \,,\nonumber\\
m_3 &=& \sqrt{\Delta m^2_{31}} = \sqrt{\frac{33}{2} \left(13+3 \sqrt{17}\right)} m_s \approx 20.46\, m_s \,. 
\end{eqnarray}
Specifically the ratio 
\begin{eqnarray}
\frac{\Delta m^2_{21}}{\Delta m^2_{31}} = \frac{m_2^2}{m_3^2} = \frac{13-3 \sqrt{17}}{13+3 \sqrt{17}} \approx 0.0247
\end{eqnarray}
is independent of the mass parameter $m_s$. 
The only difference is the Majorana phase. However, it is of little use because that phase cannot easily be accessed. Instead, we give the prediction of the effective neutrino mass parameter in neutrino-less double beta decay, which is the same in two cases, $m_{\beta\beta} = m_s$. The sum of neutrino masses is also the same, $m_1+m_2+m_3 = \sqrt{561} m_s$. The ratio of these two mass parameters is given by
\begin{eqnarray}
\frac{m_{\beta\beta}}{m_1+m_2+m_3} \approx 0.0422\,.
\end{eqnarray} 

Following \cite{King:2018kka}, it is convenient to work with the Hermitian matrix $H_\nu = M_\nu^\dag M_\nu$ instead of $M_\nu$ since $H_\nu$ preserves the $\mu-\tau$ reflection symmetry. $H_\nu$ is directly given by
\begin{eqnarray}
\frac{H_\nu}{m_s^2} = 11 \left(
\begin{array}{ccc}
 1 & -1\pm2 i \sqrt{3} & 1\pm2 i \sqrt{3} \\
 -1\mp2 i \sqrt{3} & 19 & 17+4 i \sqrt{3} \\
 1\mp2 i \sqrt{3} & 17-4 i \sqrt{3} & 19 \\
\end{array}
\right) \,
\end{eqnarray}
for case I and case II, respectively. 
They satisfy the following structure 
\begin{eqnarray}
(H_\nu)_{12} = - (H_\nu)_{13}^* \,, \quad
(H_\nu)_{22} = (H_\nu)_{33} \,,
\end{eqnarray}
from which one can directly prove $\theta_{23}=45^\circ$ and $\delta=270^\circ$. The difference of $H_\nu$ between two cases can be rotated away by redefinition of the unphysical phases in the charged lepton sector. Therefore, all oscillation parameters, including $\theta_{13}$, $\theta_{12}$, $\theta_{23}$, $\delta$, as well as mass parameters $\Delta m^2_{21}$ and $\Delta m^2_{31}$, are predicted to be exactly the same, as have been obtained in Eqs.~\eqref{eq:mixing} and \eqref{eq:mass}. Without respecting the Majorana phase and unphysical phases, the PMNS matrix in both cases takes the same form as
\begin{eqnarray}
U=
\left(
\begin{array}{ccc}
 \frac{2}{\sqrt{6}} & \frac{c_+}{\sqrt{6}} & \frac{c_-}{\sqrt{6}} \\
 \frac{1}{\sqrt{6}} & -\frac{c_+}{\sqrt{6}}-i\frac{c_-}{2} & -\frac{c_-}{\sqrt{6}}+i\frac{c_+}{2} \\
 \frac{1}{\sqrt{6}} & -\frac{c_+}{\sqrt{6}}+i\frac{c_-}{2} & -\frac{c_-}{\sqrt{6}}-i\frac{c_+}{2} \\
\end{array}
\right) \,.
\end{eqnarray}
The mixing matrix respects $\mu-\tau$ reflection symmetry and is a special case of tri-maximal TM$_1$ mixing \cite{Xing:2006ms,Albright:2008rp,Albright:2010ap,He:2011gb,Rodejohann:2012cf,Varzielas:2012pa,Grimus:2013rw}, 
with a fixed reactor angle and a fixed solar angle.

This model is not fully consistent with the oscillation data since both the predicted $\theta_{13}$ and ratio of mass square differences $\Delta m^2_{21}/\Delta m^2_{31}$ are smaller than the current global data of neutrino oscillation in $3\sigma$ ranges. 
As a comparison, current data give $\theta_{13} \sim (8.09^\circ, 8.98^\circ)$ and $\Delta m^2_{21}/\Delta m^2_{31} \sim (0.0262,0.0334)$ in $3\sigma$ ranges \cite{Esteban:2016qun}.  The explicit flavour texture of the $\mu\tau$-LSS model is corrected due to radiative corrections. We wonder if the $\mu\tau$-LSS model can be compatible with current data after the RG running effect is included. Different from \cite{King:2018kka}, where only case II is listed, here we write out both cases explicitly since radiative corrections have different contributions to $\mu$ and $\tau$ flavours.

\section{Radiative corrections to the model \label{sec:RG}}

In this section, we are going to explore how the oscillation parameters are modified by including radiative corrections.

We assume the flavour structure of the $\mu\tau$-LSS model is valid at a new scale $\Lmutau$. In order to gain a relatively large RG running effect, this scale should be sufficiently higher than the electroweak scale $\Lew$. $\Lmutau$ in principle could be different from the seesaw scale $\Lss$, but we assume they are close to each other, and thus running between $\Lmutau$ and $\Lss$ is negligible. Once heavy degrees of freedom decouple from the theory below $\Lss$, the neutrino mass and flavour mixing is governed by the dimension-5 Weinberg operator 
\begin{eqnarray}\label{eq:Weinberg}
\mathcal{L} \supset \overline{\ell} \tilde{H} \, \kappa \, \ell^c \tilde{H} + {\rm h.c.}\,,
\end{eqnarray}
where $\kappa$ is a $3 \times 3$ coupling matrix and $\tilde{H} = i\sigma_2 H^*$. After the electroweak symmetry breaking, the Higgs gains the VEV $\langle H \rangle = v_H=175$ GeV, the neutrino mass is given by $M_\nu = \kappa v_H^2$. In our following discussion, we will always denote $\kappa v_H^2$ by the effective mass matrix $M_\nu$ at any scale no matter lower or higher than the electroweak scale. For scale higher than the electroweak scale, $M_\nu$ should not be understood as neutrino masses, but just the coupling matrix with its unit normalised by $v_H^2$. 

RG running below $\Lss$ do not need to include any heavy degrees of freedom in the RG running. 
The RG running of the coupling matrix $\kappa$ was first discussed in \cite{Chankowski:1993tx,Babu:1993qv}. $M_\nu$ at two scales due to the radiative correction can be written as an integrated from as \cite{Ellis:1999my,Fritzsch:1999ee,Xing:2000ea,Mei:2003gn} 
\begin{eqnarray}
M_\nu(\Lew)=
I_\alpha\left(
\begin{array}{ccc}
 I_e & 0 & 0 \\
 0 & I_\mu & 0 \\
 0 & 0 & I_\tau
\end{array}
\right)
M_\nu(\Lmutau)
\left(
\begin{array}{ccc}
 I_e & 0 & 0 \\
 0 & I_\mu & 0 \\
 0 & 0 & I_\tau
\end{array}
\right)\,,
\label{RGE}
\end{eqnarray}
where
\begin{eqnarray}
I_\alpha&=&\text{exp}\left[-\frac{1}{16\pi^2} \int^{\text{ln}\Lmutau}_{\text{ln}\Lew} \alpha(t) \text{d}t \right]\,, \nonumber\\
I_l&=&\text{exp}\left[-\frac{C}{16\pi^2} \int^{\text{ln}\Lmutau}_{\text{ln}\Lew} y^2_l(t) \text{d}t \right]\,,
\end{eqnarray}
for $l=e,\mu,\tau$. Here we have ignored the difference between $M_\nu$ at $\Lmutau$ and that just below $\Lss$.
In the SM and the minimal supersymmetric model (MSSM), $C$ and $\alpha$ are given by 
\begin{eqnarray}
&&C_\text{SM}=-\frac{3}{2}\,, \qquad \alpha_\text{SM}\approx-3g^2_2+\lambda+6y^2_t\,, \nonumber\\
&&C_\text{MSSM}=1\,, \qquad \alpha_\text{MSSM}\approx-\frac{6}{5}g^2_1-6g^2_2+6y^2_t\,,
\label{RGcoefficients}
\end{eqnarray}
respectively, where $g_{1,2}$ denote the gauge couplings, $\lambda$ denotes the quartic Higgs coupling in the SM, and $y_t$, $y_l$ (for $l=e,\mu,\tau$) are Yukawa couplings of the top quark and charged leptons, respectively. In MSSM, the Higgs $\tilde{H}$ contributing to the Weinberg operator in Eq.~\eqref{eq:Weinberg} should be replaced by $H_u$, and the VEV $v_H$ contributing to the neutrino mass $M_\nu$ should be replaced by $v_{H_u}=v_H \sin\beta$. 

We see that in Eq.~\eqref{RGE}, $I_\alpha$ is an overall factor affecting the magnitudes of the absolute neutrino masses, and $I_l$ are flavour-dependent corrections which may modify the mass structure and flavour mixing. Due to the different signs of $C$ in SM and MSSM (c.f. Eq.~\eqref{RGcoefficients}), the flavour-dependent corrections go to opposite directions in the SM and MSSM. We follow the approximation proposed in \cite{Zhou:2014sya}: the Yukawa couplings $y_e$, $y_\mu$ are too small as compared with $y_\tau$ such that thus $I_e$ and $I_\mu$ can be approximately set to be identities, and $I_\tau$ is re-parametrised as $1+\epsilon$, where
\begin{eqnarray}
\epsilon&=&I_\tau-1 \approx -\frac{C}{16\pi^2} \int^{\text{ln}\Lambda_{\mu\tau}}_{\text{ln}\Lambda_\text{EW}} y^2_\tau(t) \text{d}t\,. 
\end{eqnarray}
In the case of slowing varying Yukawa coupling, $y_\tau(t)$ can be replaced by $y_{\tau, \text{EW}} = m_\tau / v_H$ in SM (or $m_\tau / (v_H \sin\beta)$ in MSSM) and $\epsilon$ is approximated to $\epsilon \approx-\frac{C}{16\pi^2} y^2_{\tau, \text{EW}} \text{ln}\frac{\Lambda_{\mu\tau}}{\Lambda_\text{EW}}$ with $y_{\tau, \text{EW}}$ being the $\tau$-lepton Yukawa coupling at the electroweak scale. Since $C$ is negative in SM (positive in MSSM), the correction $\epsilon$ is positive in SM (negative in MSSM). 

At the scale $\Lmutau$, $M_\nu(\Lmutau)$ takes the exact form as in Eq.~\eqref{eq:mutau_mass}. 
With the help of the above approximation, the Majorana mass matrix at the electroweak scale is represented by
\begin{eqnarray}\label{eq:mutau_mass2}
\text{Case I: } & M_\nu(\Lambda_{\rm EW})=\tilde{m}_s \left(
\begin{array}{ccc} \label{eq:mass_RG}
 1 & 3 & 1+\epsilon \\
 3 & 9+11 \omega  & (3+11 \omega)(1+\epsilon)  \\
 1+\epsilon & (3+11 \omega)(1+\epsilon)  & (1+11 \omega) (1+\epsilon)^2 \\
\end{array}
\right) \,, \nonumber\\
\text{Case II: } & M_\nu(\Lambda_{\rm EW})=\tilde{m}_s \left(
\begin{array}{ccc}
 1 & 1 & 3(1+\epsilon) \\
 1 & 1+11 \omega ^2 & (3+11 \omega ^2)(1+\epsilon) \\
 3(1+\epsilon) & (3+11 \omega ^2)(1+\epsilon) & (9+11 \omega ^2)(1+\epsilon)^2 \\
\end{array}
\right) \,,
\end{eqnarray} 
where $\tilde{m}_s = I_\alpha m_s$. Only two real parameters are involved in the mass matrix $M_\nu$ at the electroweak scale. One of them, $\tilde{m}_s$ contributes only to the absolute values of neutrino masses. The exact value of $m_s$ or $I_\alpha$ is not important at low energy theory. Only their combination $\tilde{m}_s$ is important. The other parameter $\epsilon$, representing the RG running effect, is the only parameter contributing to flavour mixing and the ratio of mass square differences. It also violates the $\mu\tau$ conjugation relation between the two mass matrices. 

The Hermitian matrix $H_\nu$ at the electroweak scale is directly obtained from Eq.~\eqref{eq:mass_RG}. In order to get the analytical approximate results of the oscillation parameters, we expand $H_\nu$ in order of $\epsilon$ as   
\begin{eqnarray} \label{eq:H_RG}
\frac{H_\nu(\Lew)}{\tilde{m}_s^2} = \frac{H_\nu(\Lmutau)}{m_s^2} + \frac{\delta H_\nu}{\tilde{m}_s^2}
\end{eqnarray}
with
\begin{eqnarray}
\frac{\delta H_\nu}{\tilde{m}_s^2} &=& 
\left(
\begin{array}{ccc}
 2 & -5+11 i \sqrt{3} & -20+33 i \sqrt{3} \\
 -5-11 i \sqrt{3} & 194 & 391+66 i \sqrt{3} \\
 -20-33 i \sqrt{3} & 391-66 i \sqrt{3} & 640 \\
\end{array}
\right) \epsilon + \mathcal{O}(\epsilon^2) \,, \nonumber\\
\frac{\delta H_\nu}{\tilde{m}_s^2} &=& 
\left(
\begin{array}{ccc}
 18 & -15-33 i \sqrt{3} & 32-55 i \sqrt{3} \\
 -15+33 i \sqrt{3} & 194 & 351+110 i \sqrt{3} \\
 32+55 i \sqrt{3} & 351-110 i \sqrt{3} & 624 \\
\end{array}
\right) \epsilon + \mathcal{O}(\epsilon^2) 
\end{eqnarray}
for case I and case II, respectively. On the right hand side of Eq.~\eqref{eq:H_RG}, only one free parameter $\epsilon$ appears. The $\mu-\tau$ reflection symmetry is not preserved any more. 
The RG running effect specifies the $\tau$ sector, and thus two cases in Eq.~\eqref{eq:mutau_mass} gain totally different corrections. 

By perturbatively diagonalising $H_\nu$, we obtain corrections to both $\theta_{13}$ and the ratio of mass square differences $\Delta m^2_{21}/\Delta m^2_{31}$, which are determined by $\epsilon$. 
Including the other parameters, the corrected oscillation parameters are approximatively given by
\begin{eqnarray} \label{eq:prediction}
\theta_{13} &\approx& 7.807^\circ - 8.000^\circ \epsilon \,, \quad \nonumber\\
\theta_{12} &\approx& 34.50^\circ -12.30^\circ \epsilon \,, \quad \nonumber\\
\theta_{23} &\approx& 45.00^\circ -31.64^\circ \epsilon \,, \quad \nonumber\\
\delta &\approx& 270.00^\circ + 3.23^\circ \epsilon\,, \quad \nonumber\\
\frac{\Delta m^2_{21}}{\Delta m^2_{31}} &\approx& 0.0247 - 0.0147 \epsilon
\end{eqnarray} 
in case I, and
\begin{eqnarray} \label{eq:prediction_caseII}
\theta_{13} &\approx& 7.807^\circ + 0.345^\circ \epsilon \,, \quad \nonumber\\
\theta_{12} &\approx& 34.50^\circ -13.96^\circ \epsilon \,, \quad \nonumber\\
\theta_{23} &\approx& 45.00^\circ -30.50^\circ \epsilon \,, \quad \nonumber\\
\delta &\approx& 270.00^\circ + 2.33^\circ \epsilon\,, \quad \nonumber\\
\frac{\Delta m^2_{21}}{\Delta m^2_{31}} &\approx& 0.0247 - 0.0249 \epsilon
\end{eqnarray}
in case II. Here again, $I_\alpha$ gives only an overall enhancement or suppression to masses and thus does not contribute to the above formulas.

Let us first have a look at case II. This case is not compatible with data after the RG running is included. Reasons are given below. 
In Eq.~\eqref{eq:prediction_caseII}, we can see that $\theta_{13}$ gains a very small correction from $\epsilon$. In order to enhance $\theta_{13}$ by $0.2^\circ$, $\epsilon$ should be positive and not smaller than $0.5$, in spite of validity of perturbation calculation. In MSSM, $\epsilon$ is always a negative parameter and thus, does not satisfy the requirement. In the SM, $\epsilon$ is positive, but the induced correction is too small. Furthermore, $\theta_{13}$ and ${\Delta m^2_{21}}/{\Delta m^2_{31}}$ always gain corrections in opposite directions. If one parameter runs closer to the experimental allowed range, the other runs farther away. Therefore, Eq.~\eqref{eq:prediction_caseII} is not consistent with current oscillation data. In this work, we have assumed $\Lmutau$ close to the seesaw scale $\Lss$. If such an assumption is given up, e.g., $\Lmutau \gg \Lss$, heavy neutrinos may contribute to the running effect from $\Lmutau$ to $\Lss$, the RG running behaviour could be modified, and case II may be still allowed by data. We will not consider this possibility in our paper.

\begin{figure}
\centering
\includegraphics[scale=1]{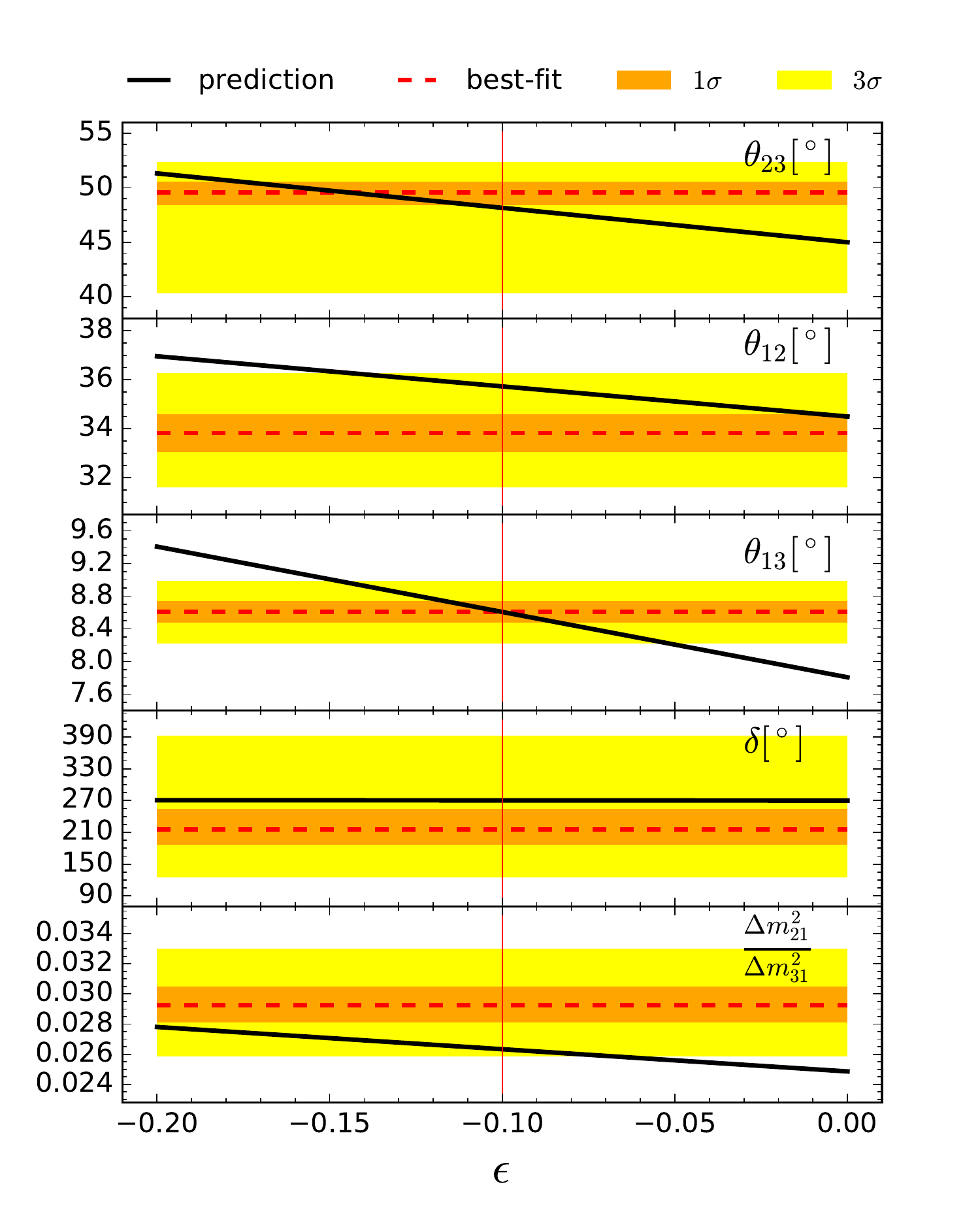}
\caption{\label{fig:oscillation_parameters} Oscillation parameters modified by radiative corrections in case I of the $\mu\tau$-LSS model. Best-fit values, $1\sigma$ and 3$\sigma$ ranges of global fits data of neutrino oscillation experiments are shown as comparisons. }
\end{figure}

Then, we turn back to case I. Oscillation parameters as functions of the RG running parameter $\epsilon$ are shown in Fig. \ref{fig:oscillation_parameters}. In this case, all parameters can be compatible with current oscillation data in $3\sigma$ ranges with a suitable value for the RG running parameter $\epsilon$. Specifically, both $\theta_{13}$ and ${\Delta m^2_{21}}/{\Delta m^2_{31}}$ are corrected in the same direction. To increase their values, $\epsilon$ has to be negative with value $-\epsilon \sim \mathcal{O}(0.1)$. However, these two parameters cannot be compatible with each other in $1\sigma$ ranges. We have compared the linear approximation in Eq.~\eqref{eq:prediction} with the full one-loop RG running code in MSSM \cite{Mei:2003gn,Zhou:2014sya} and confirm that it is valid to a very high precision level. 
By setting $\Lew$ and $\Lambda_{\mu\tau}$ around $10^2$ and $10^{14}$ GeV, respectively, we vary $\tan\beta$ in $(0,100)$ and obtain correlation between $\tan\beta$ and $\epsilon$. The value of $\epsilon$ of order 0.1 refers to a large $\tan\beta$, e.g., $\epsilon = 0.05,  0.1$ corresponding to $\tan\beta \sim 40, 66$, respectively.  To summarise, the $\mu\tau$-LSS model with RG correction in MSSM with large $\tan\beta$ is compatible with current oscillation data.

\section{A concrete $\mu\tau$-LSS model in $S_{4} \times S_{4}$ \label{sec:model}}

In this section, we present a concrete flavour model to realise the $\mu\tau$-LSS flavour structure in case I. 
We assume the flavour symmetry to be $S_{4L} \times S_{4R}$ in the SUSY framework. How leptons gain flavoured masses based on specified flavon vacua will be discussed in section~\ref{sec:mass} and how flavons gain the specified VEVs will be given in section~\ref{sec:vacuum}. In addition, we introduce two $U(1)$'s to achieve diagonal and hierarchical masses for charged leptons and forbid unnecessary superpotential terms. An example of charge assignment in $S_{4L} \times S_{4R} \times U(1) \times U(1)'$ symmetries is given in section~\ref{sec:charge}.

\subsection{Fermion masses \label{sec:mass}}

Three left-handed $SU(2)_L$ doublets of leptons $\ell$ transform as a triplet in $S_{4L}$ but a trivial singlet in $S_{4R}$. We introduce two right-handed neutrinos $N_1$ and $N_2$, transforming as a doublet in $S_{4R}$ and blind in $S_{4L}$. In order to realise the flavour structure, we introduce four flavon multiplets $\phi_t'$, $\phi_N$, $\xi_{TS}$ and $\xi'_t$. The flavon $\phi_N$ talks to both left-handed and right-handed fermions, and thus transforms non-trivially as a triplet-doublet $(\mathbf{3}_L, \mathbf{2}_R)$ in $S_{4L} \times S_{4R}$, where the subscripts $_L$ and $_R$ specifying groups $S_{4L}$ and $S_{4R}$, respectively. The other flavons, $\phi_t'$ interacts with the SM leptons, arranged as a triplet $\mathbf{3}'_L$ of $S_{4L}$ and non-trivial singlet $\mathbf{1}'_R$ of $S_{4R}$. $\xi_{TS}$ and $\xi'_t$ only interact with right-handed neutrinos, arranged as triplets $\mathbf{3}_R$, $\mathbf{3}'_R$ of $S_{4R}$, respectively. These representation arrangements are simply summarised as
\begin{eqnarray} \label{eq:fields}
&&\ell = \begin{pmatrix} \ell_1 \\ \ell_2 \\ \ell_3 \end{pmatrix} \sim (\mathbf{3}_L,\mathbf{1}_R) \,, \quad
\phi'_t = \begin{pmatrix} \phi'_{t,1} \\ \phi'_{t,2} \\ \phi'_{t,3} \end{pmatrix} \sim (\mathbf{3}'_L,\mathbf{1}'_R) \,, \nonumber\\
&&
e^c\sim (\mathbf{1}'_L, \mathbf{1}'_R)\,,\quad
\mu^c\sim (\mathbf{1}_L, \mathbf{1}_R)\,,\quad
\tau^c\sim (\mathbf{1}'_L, \mathbf{1}'_R)\,, \nonumber\\
&&N = (N_1, N_2) \sim (\mathbf{1}_L, \mathbf{2}_R) \,, \nonumber\\
&&\phi_N \equiv (\phi_{\rm atm}, \phi_{\rm sol}) = \begin{pmatrix} \phi_{\rm atm,1} & \phi_{\rm sol,1} \\ \phi_{\rm atm,2} & \phi_{\rm sol,2} \\ \phi_{\rm atm,3} & \phi_{\rm sol,3} \end{pmatrix} \sim (\mathbf{3}_L, \mathbf{2}_R) \,, \nonumber\\
&&\xi_{TS} = (\xi_{TS,1}, \xi_{TS,2}, \xi_{TS,3}) \sim (\mathbf{1}_L, \mathbf{3}_R) \,, \quad
\xi'_t = (\xi'_{t,1}, \xi'_{t,2}, \xi'_{t,3}) \sim (\mathbf{1}_L, \mathbf{3}'_R)\,.
\end{eqnarray}
We make the convention that $S_{4L}$ acts on multiplets vertically and $S_{4R}$ acts horizontally. 
Vacuum alignments are assumed (and later justified) to be
\begin{eqnarray} \label{eq:flavon_VEV}
\langle \phi'_t \rangle &=& \begin{pmatrix} 0 \\ 1 \\ 0 \end{pmatrix} v_{\phi'_t} \,, \nonumber\\
\langle \phi_N \rangle &=& \begin{pmatrix} 0 & \frac{1}{\sqrt{11}} \\ \frac{1}{\sqrt{2}} & \frac{3}{\sqrt{11}} \\ \frac{-1}{\sqrt{2}} & \frac{-1}{\sqrt{11}} \end{pmatrix} v_{\phi_N} \,, \nonumber\\
\langle \xi_{TS} \rangle &=& (\frac{1}{3},-\frac{2}{3} \omega,-\frac{2}{3} \omega^2) v_{\xi_{TS}} \,,\nonumber\\
\langle \xi'_t \rangle &=& (0,1,0) v_{\xi'_t} \,.
\end{eqnarray}
Note that $\phi'_t$, $\phi_N$, $\xi_{TS}$, $\xi'_t$ and $\ell$, $N^c$ are not the only particles introduced in the model. More flavons, as well as driving fields, have to be introduced to achieve the vacuum alignment self-consistently, and  will be discussed in detail in the next subsection. 

Terms for generating charged lepton and neutrino masses are given by
\begin{eqnarray} \label{eq:superpotential_leptons}
&&w_\ell = \frac{y_e}{\Lambda^3} \ell (\phi'_t)^3 e^c H_d +
\frac{y_\mu}{\Lambda^2} \ell (\phi'_t)^2 \mu^c H_d +
\frac{y_\tau}{\Lambda} \ell \phi'_t \tau^c H_d \,,\nonumber\\
&&w_\nu = \frac{y}{\Lambda} \ell H_u \phi_N N^c + \frac{\lambda}{\Lambda} (N^c N^c)_{\mathbf{2}_R} (\xi_{TS} \xi'_t)_{\mathbf{2}_R} \,,
\end{eqnarray}
where $
(\phi'_t)^3 = a_1 (\phi'_t \phi'_t)_{(\mathbf{1}_L, \mathbf{1}_R)} \phi'_t + a_2 ((\phi'_t \phi'_t)_{(\mathbf{2}_L, \mathbf{1}_R)} \phi'_t)_{(\mathbf{3}'_L, \mathbf{1}'_R)} + a_3 ((\phi'_t \phi'_t)_{(\mathbf{3}_L, \mathbf{1}_R)} \phi'_t)_{(\mathbf{3}'_L, \mathbf{1}'_R)}$ represents any $(\mathbf{3}'_L, \mathbf{1}_R)$ contractions of trilinear couplings of $\phi'_t$ with $a_{1,2,3}$ being dimensionless coefficients, 
$(\phi'_t)^2 = (\phi'_t \phi'_t)_{(\mathbf{3}_L, \mathbf{1}_R)} $ is a bilinear $(\mathbf{3}_L, \mathbf{1}_R)$ contractions of $\phi'_t$, and
$H_u$ is a trivial singlet in both $S_{4L}$ and $S_{4R}$. After the $\phi'_t$ gains the VEV, we arrive at
\begin{eqnarray}
\langle (\phi'_t)^3 \rangle = (a_2-2a_3) v_{\phi'_t}^3 \begin{pmatrix} 1 \\ 0 \\ 0 \end{pmatrix} \,,~~
\langle (\phi'_t)^2 \rangle = v_{\phi'_t}^2 \begin{pmatrix} 0 \\ 0 \\ 1 \end{pmatrix} \,,~~
\langle (\phi'_t) \rangle = v_{\phi'_t} \begin{pmatrix} 0 \\ 1 \\ 0 \end{pmatrix} \,,~~
\end{eqnarray}
Its VEV, as well as the Higgs VEV, results in diagonal charged lepton mass matrix with diagonal entries given by
\begin{eqnarray}
m_e= y_e (a_2-2a_3) \frac{v_{\phi'_t}^3}{\Lambda^3} v_{H_d} \,,~~
m_\mu = y_\mu \frac{v_{\phi'_t}^2}{\Lambda^2} v_{H_d} \,,~~
m_\tau = y_\tau \frac{v_{\phi'_t}}{\Lambda} v_{H_d} \,.
\end{eqnarray}
Below, we will focus on mass matrices in the neutrino sector.

Based on the $\phi_N$ vacuum alignment, we obtain the Dirac mass matrix as
\begin{eqnarray}
M_D= y \begin{pmatrix} \frac{1}{\sqrt{11}} & 0 \\ \frac{-1}{\sqrt{11}} & \frac{-1}{\sqrt{2}} \\ \frac{3}{\sqrt{11}} & \frac{1}{\sqrt{2}} \end{pmatrix} \frac{v_{\phi_N}}{\Lambda} v_{H_u}. 
\end{eqnarray} 
Here, the minus sign in the last row of $M_D$ is unphysical, which can be absorbed by re-defining $\ell_\tau \to -\ell_\tau$. 
The doublet contraction of $\xi_{TS}$ and $\xi'_t$ gives rise to 
\begin{eqnarray}
\langle (\xi_{TS} \xi'_t)_{\mathbf{2}_R} \rangle = (2\omega,1) \frac{-v_{\xi_{TS}} v_{\xi'_t}}{3} \,,
\end{eqnarray}
leading to the Majorana mass matrix 
\begin{eqnarray}
M_M = \lambda \begin{pmatrix} 2\omega & 0 \\ 0 & 1 \end{pmatrix} \frac{-v_{\xi_{TS}} v_{\xi'_t}}{3\Lambda} \,.
\end{eqnarray}
After right-handed neutrinos are integrated out, according to the seesaw mechanism, the active neutrino coupling matrix is given by
\begin{eqnarray}
M_\nu = m_s \begin{pmatrix}
 1 & 3 & 1 \\
 3 & 9+11 \omega  & 3+11 \omega  \\
 1 & 3+11 \omega  & 1+11 \omega  \\
\end{pmatrix}
\end{eqnarray}
which reproduces Case I of Eq.\ref{eq:mutau_mass},
with 
\begin{eqnarray}
m_s = \frac{3 \omega^2 y^2 v_{\phi_N}^2 v_{H_u}^2}{22 \lambda v_{\xi_T} v_{\xi_t} \Lambda} \,.
\end{eqnarray}

\subsection{Vacuum alignments\label{sec:vacuum}}

The vacuum alignment, in particular for $\phi_N$ in Eq.\ref{eq:flavon_VEV}, is not obvious. We set up this subsection for a detailed analysis of how these flavons gain the required VEVs. More flavons, which do not directly contribute to lepton masses, but influence on the other flavon VEVs, have to be introduced. 
First of all, we introduce three flavons $\phi_T\sim (\mathbf{3}_L, \mathbf{1}_R)$, $\phi_S\sim (\mathbf{3}_L, \mathbf{1}_R)$, $\phi_U\sim (\mathbf{3}'_L, \mathbf{1}_R)$
and require their VEVs invariant under the transformation of generators $T$, $S$, $U$ of $S_4$, respectively, 
\begin{eqnarray}\label{eq:VEV_STU}
\langle \phi_T \rangle = \begin{pmatrix} 1\\0\\0 \end{pmatrix} v_{\phi_T}\,,\quad
\langle \phi_S \rangle = \begin{pmatrix} 1\\1\\1 \end{pmatrix} v_{\phi_S}\,,\quad
\langle \phi_U \rangle = \begin{pmatrix} 0\\1\\-1 \end{pmatrix} v_{\phi_U}\,.
\end{eqnarray}
The generators $S$, $T$ and $U$ are given in appendix~\ref{app:S4}.
These VEVs can be easily obtained and have been discussed in a lot of $S_4$ models (see e.g., \cite{Hagedorn:2010th,Hagedorn:2012ut,Ding:2013hpa,Feruglio:2013hia}). Here, we give an example, with driving terms for these VEVs given by
\begin{eqnarray} \label{eq:superpotential_driving1}
w_d \supset \phi^d_T (\phi_T\phi_T)_{(\mathbf{2}_L, \mathbf{1}_R)} + 
\phi^d_S (\phi_S\phi_S)_{(\mathbf{3}_L, \mathbf{1}_R)} +
\phi^d_U \left[A_U \phi_U + (\phi_T\phi_S)_{(\mathbf{3}'_L, \mathbf{1}_R)} \right] \,,
\end{eqnarray}
Here and in the following, we only consider renormalisable couplings. And any dimensionless coefficients which do not influence to our later discussion are ignored. $A_U$ is a normalised parameter with a mass unit. The driving fields are arranged as suitable multiplets to keep each term satisifying the flavour symmetry. 
Minimisation of the superpotential respect to the driving fields $\phi^d_T$, $\phi^d_S$ and $\phi^d_U$ gives the following equations 
\begin{eqnarray}
&&(\phi_T\phi_T)_{(\mathbf{2}_L, \mathbf{1}_R)} = \begin{pmatrix}\phi_{T,2}^2 + 2\phi_{T,1} \phi_{T,3} \\ \phi_{T,3}^2 + 2\phi_{T,1} \phi_{T,2} \end{pmatrix} = 0 \,, \nonumber\\
&&(\phi_S\phi_S)_{(\mathbf{3}_L, \mathbf{1}_R)} = 2 \begin{pmatrix}\phi_{S,1}^2 -\phi_{S,2} \phi_{S,3} \\ \phi_{S,3}^2 - \phi_{S,1} \phi_{S,2} \\ \phi_{S,2}^2 -\phi_{S,1} \phi_{S,3} \end{pmatrix} = 0 \,, \nonumber\\
&&A_U \phi_U + (\phi_T\phi_S)_{(\mathbf{3}'_L, \mathbf{1}_R)} = A_U \begin{pmatrix}\phi_{U,1} \\ \phi_{U,2} \\ \phi_{U,3} \end{pmatrix} + 
\begin{pmatrix} \phi_{T,2} \phi_{S,3} - \phi_{T,3} \phi_{S,2} \\ \phi_{T,1} \phi_{S,2} - \phi_{T,2} \phi_{S,1} \\ \phi_{T,3} \phi_{S,1} - \phi_{T,1} \phi_{S,3} \end{pmatrix}= 0 \,,
\end{eqnarray}
respectively. The first two equations determine directions of $\langle \phi_T \rangle$ and $\langle \phi_S \rangle$ in Eq.~\eqref{eq:VEV_STU} with $v_{\phi_T}$ and $v_{\phi_S}$ undetermined. Taking $\langle \phi_T \rangle$ and $\langle \phi_S \rangle$ to the third equation $\langle \phi_T \rangle$ is determined with correlation $v_{\phi_U} = -v_{\phi_T} v_{\phi_S}/A_{U}$ satisfied. 

It is worth noting that full solutions for $(\phi_T\phi_T)_{(\mathbf{2}_L, \mathbf{1}_R)} =0$ are given by 
\begin{eqnarray}
(1,0,0)^T v_{\phi_T}\,,\quad
(\frac{1}{3},-\frac{2}{3},-\frac{2}{3})^T v_{\phi_T}\,,\quad
(\frac{1}{3},-\frac{2}{3}\omega,-\frac{2}{3}\omega^2)^T v_{\phi_T}\,,\quad
(\frac{1}{3},-\frac{2}{3}\omega^2,-\frac{2}{3}\omega)^T v_{\phi_T}
\end{eqnarray} 
with $v_{\phi_T}$ undetermined. All these solution are related with each other by $S_{4L}$ conjugacy transformation. By randomly choosing one of these solutions as the VEV, one can always rotate it into the $(1,0,0)^T$ direction  (see Appendix~\ref{app:degeneracy}). Therefore, we fix the flavon VEV at the first solution without loss of generality. 

The VEV of $\phi'_t$ can be obtained by evolving $\phi_T$ in its driving terms
\begin{eqnarray}\label{eq:superpotential_driving2}
w_d \supset \phi^d_t (\phi'_t\phi'_t)_{(\mathbf{1}_L, \mathbf{1}_R)} + \phi^{d\prime}_t (\phi_T\phi'_t)_{(\mathbf{1}'_L, \mathbf{1}_R)} \,.
\end{eqnarray}
Taking $\langle \phi_T \rangle$ into account, minimisation of these terms is explicitly written out as
\begin{eqnarray}
&&(\phi'_t\phi'_t)_{(\mathbf{1}_L, \mathbf{1}_R)}|_{\langle \phi_T \rangle} = \phi_{t,1}^{\prime2} + 2 \phi'_{t,2} \phi'_{t,3} = 0 \,, \nonumber\\ 
&&(\phi_T\phi'_t)_{(\mathbf{1}'_L, \mathbf{1}_R)}|_{\langle \phi_T \rangle} = v_{\phi_T} \phi'_{t,1} = 0 \,,
\end{eqnarray}
which leads to $\phi'_{t,1} = 0$ and $\phi'_{t,2} \phi'_{t,3} =0$. Without lose of generality, we choose $\phi'_{t,3}=0$ and $\phi'_{t,2} = v_{\phi'_t}$ with $v_{\phi'_t}$ undetermined. 
We introduce another triplet flavon $\tilde{\phi}_t'$ for our later use. With similar constructions of the driving terms as in Eq.~\eqref{eq:superpotential_driving2}, 
\begin{eqnarray}\label{eq:superpotential_driving3}
w_d \supset \tilde{\phi}^d_t (\tilde{\phi}'_t\tilde{\phi}'_t)_{(\mathbf{1}_L, \mathbf{1}_R)} + \tilde{\phi}^{d\prime}_t (\phi_T\tilde{\phi}'_t)_{(\mathbf{1}'_L, \mathbf{1}_R)}
\end{eqnarray}
and adding one more term 
\begin{eqnarray} \label{eq:superpotential_driving4}
w_d \supset \varphi^{d}_T \left[ \mu_x^2 + (\phi'_t \tilde{\phi}'_t)_{(\mathbf{1}_L,\mathbf{1}_R)} \right] \,,
\end{eqnarray}
$\langle \tilde{\phi}'_t \rangle = (0,0,1)^T v_{\tilde{\phi}'_t}$ can be determined and the correlation $\langle \mu_x^2 \rangle+v_{\phi'_t}v_{\tilde{\phi}'_t}=0$ is obtained. Here, $\mu_x^2$ is not a free parameter but a contraction of some other flavons. Its exact expression will be given later after the charge assignment is complete. 

We then consider flavons which transform non-trivially in $S_{4R}$. We introduced additional $\xi_T$. This flavon, together with $\xi_{TS}$ and $\xi'_t$, are arranged as $(\mathbf{1}_L, \mathbf{3}_R)$, $(\mathbf{1}_L, \mathbf{3}_R)$ and $(\mathbf{1}_L, \mathbf{3}'_R)$, respectively. The driving terms are given by
\begin{eqnarray}\label{eq:superpotential_driving5}
w_d \supset \xi^d_T (\xi_T\xi_T)_{(\mathbf{1}_L, \mathbf{2}_R)} + \xi^d_{TS} (\xi_{TS}\xi_{TS})_{(\mathbf{1}_L, \mathbf{2}_R)} + \xi^d_t (\xi'_t\xi'_t)_{(\mathbf{1}_L, \mathbf{1}_R)} + \xi^{d\prime}_t (\xi_T\xi'_t)_{(\mathbf{1}_L, \mathbf{1}'_R)}\,. 
\end{eqnarray}
Minimisation of the first two terms lead to $(\xi_T\xi_T)_{(\mathbf{1}_L, \mathbf{2}_R)} = (\xi_{TS}\xi_{TS})_{(\mathbf{1}_L, \mathbf{2}_R)} =0$. Full solutions for $\xi_T$ are given by 
\begin{eqnarray}
(1,0,0) v_{\xi_T}\,,\quad
(\frac{1}{3},-\frac{2}{3},-\frac{2}{3}) v_{\xi_T}\,,\quad
(\frac{1}{3},-\frac{2}{3}\omega,-\frac{2}{3}\omega^2) v_{\xi_T}\,,\quad
(\frac{1}{3},-\frac{2}{3}\omega^2,-\frac{2}{3}\omega) v_{\xi_T}
\end{eqnarray} 
with $v_{\xi_T}$ undetermined. Those for $\xi_{TS}$ can be similarly written out.
VEVs of $\xi_T$ and $\xi_{TS}$ could be any of them, respectively. In the case that both flavons preserve $Z_3$ symmetries, there is a larger probability that the direction $\langle \xi_{TS} \rangle$ is different from that of $\langle \xi_{T} \rangle$. And therefore, the $S_4$ transformation cannot rotate both directions to $(1,0,0)$. Instead, we can fix $\langle \xi_T \rangle$ at  $(1,0,0)v_{\xi_T}$, and $\xi_{TS}$ at $(\frac{1}{3},-\frac{2}{3}\omega,-\frac{2}{3}\omega^2) v_{\xi_{TS}}$ as in Eq.~\eqref{eq:flavon_VEV}. The later is invariant under a different $Z_3$ symmetry generated by $TS$. For more detail of how to determine these VEVs, please see Appendix~\ref{app:degeneracy}. Note that our model with current setup cannot fully determine the $\xi_{TS}$ VEV, but leaves a large possibility for $\xi_{TS}$ to take such a required VEV. In order to determine the $\xi_{TS}$ VEV, another way could be to consider cross couplings between $\xi_{TS}$ and $\xi_T$. With suitable small cross couplings between $\xi_{TS}$ and $\xi_T$, the global vacuum may be obtained when $\langle \xi_{TS}\rangle $ and $\langle \xi_T \rangle$ take different directions, and this vacuum degeneracy may be be avoided. 
The last two driving terms determine the VEV $\langle \xi'_t \rangle$. 
Once $\langle \xi_T \rangle$ is fixed to be $\propto(1,0,0)$, we derive $\langle \xi'_t \rangle = (0,1,0) v_{\xi'_t}$, following similar discussion as that for $\phi'_t$. 


To achieve the VEV for $\phi_N$ is a non-trivial task. Let us first denote $\phi_N$ by $\phi_N = (\phi_{\rm atm}, \phi_{\rm sol})$. Both $\phi_{\rm atm}$ and $\phi_{\rm sol}$ are triplets $\mathbf{3}_L$ of $S_{4L}$, while $\phi_{\rm atm}$ and $\phi_{\rm sol}$ form a doublet $\mathbf{2}_R$ of $S_{4R}$. For convenience, we denote VEVs of $\phi_{\rm atm}$ and $\phi_{\rm sol}$ respectively as
\begin{eqnarray} \label{eq:VEV12}
\langle \phi_{\rm atm}\rangle = \begin{pmatrix} 0 \\ \frac{-1}{\sqrt{2}} \\ \frac{1}{\sqrt{2}} \end{pmatrix} v_{\phi_{\rm atm}} \,, \quad
\langle \phi_{\rm sol} \rangle = \begin{pmatrix} \frac{1}{\sqrt{11}} \\ \frac{-1}{\sqrt{11}} \\ \frac{3}{\sqrt{11}} \end{pmatrix} v_{\phi_{\rm sol}} \,.
\end{eqnarray}
With this notation, we now address question of how to obtain the required $\phi_N$ VEV into three steps: 
\begin{itemize}

\item[I] To construct superpetential terms to separate $\phi_{\rm atm}$ and $\phi_{\rm sol}$ from the same doublet of $S_{4R}$. 
\item[II] To drive $\phi_{\rm atm}$ and $\phi_{\rm sol}$ separately to gain VEVs with different directions, i.e., one in $(0,-1,1)^T$ and the other in $(1,-1,3)$. 

\item[III] To force $v_{\phi_{\rm atm}}$ identical to $v_{\phi_{\rm sol}}$, i.e., $v_{\phi_{\rm atm}} = v_{\phi_{\rm sol}} \equiv v_{\phi_N}$. 

\end{itemize}
How to achieve each step is given as following. 

For the first step, as $\phi_{\rm atm}$ and $\phi_{\rm sol}$ form a $\mathbf{2}_R$ of $S_{4R}$, we need to take care of the correlation between directions of the $\phi_{\rm atm}$ VEV and $\phi_{\rm sol}$ VEV. One way to separate them is introducing two flavons $\rho$ and $\tilde{\rho}$, which are doublets of $S_{4R}$ and gain VEVs $\propto (1,0)$ and $(0,1)$, respectively. The singlet contraction in $S_{4R}$ leaves $\phi_{\rm sol}$ and $\phi_{\rm atm}$ separately. For the convenience of step II, we arrange these flavons also as doublets of $S_{4L}$, i.e., $\rho \sim \tilde{\rho} \sim (\mathbf{2}_L, \mathbf{2}_R)$ and their VEVs in the following form 
\begin{eqnarray}\label{eq:VEV_rho}
\langle \rho \rangle = \begin{pmatrix} 1 & 0 \\ 1 & 0 \end{pmatrix} v_\rho\,,\quad
\langle \tilde{\rho} \rangle = \begin{pmatrix} 0 & 1 \\ 0 & 1 \end{pmatrix} v_{\tilde{\rho}} \,.
\end{eqnarray} 
In $S_{4L}$, these VEVs take the direction $(1,1)^T$, invariant under the generator $U$. This is prepared for our later use in step II. In order to realise these VEVs, we construct the driving terms as
\begin{eqnarray} \label{eq:superpotential_driving6}
\hspace{-5mm}
w_d \supset \rho^d (\rho\rho)_{(\mathbf{2}_L,\mathbf{1}_R)} \!+\! 
\tilde{\rho}^d (\tilde{\rho} \tilde{\rho})_{(\mathbf{2}_L,\mathbf{1}_R)} \!+\!
\rho^{d\prime} \left[ (\rho\tilde{\rho})_{(\mathbf{2}_L,\mathbf{1}_R)} \!+\! g_U (\phi_U \phi_U)_{(\mathbf{2}_L, \mathbf{1}_R)} \right] \!+\!
\rho^{d\prime\prime} (\rho\tilde{\rho})_{(\mathbf{1}'_L,\mathbf{1}_R)} \,,
\end{eqnarray}
where $g_U$ corresponds the ratio of coefficients between $(\rho\tilde{\rho})_{(\mathbf{2}_L,\mathbf{1}_R)}$  and $(\phi_U \phi_U)_{(\mathbf{2}_L, \mathbf{1}_R)}$ terms. 
Minimisation of these terms gives rise to 
\begin{eqnarray}
&&(\rho\rho)_{(\mathbf{2}_L,\mathbf{1}_R)} = \begin{pmatrix} 2 \rho_{21} \rho_{22} \\ 2 \rho_{11} \rho_{12} \end{pmatrix} =0 \,,
\nonumber\\
&&(\tilde{\rho}\tilde{\rho})_{(\mathbf{2}_L,\mathbf{1}_R)} = \begin{pmatrix} 2 \tilde{\rho}_{21} \tilde{\rho}_{22} \\ 2 \tilde{\rho}_{11} \tilde{\rho}_{12} \end{pmatrix} =0 \,,
\nonumber\\
&&(\rho\tilde{\rho})_{(\mathbf{2}_L,\mathbf{1}_R)} + (\phi_U \phi_U)_{(\mathbf{2}_L, \mathbf{1}_R)}\big|_{\langle \phi_U \rangle} = \begin{pmatrix} \rho_{21} \tilde{\rho}_{22} + \rho_{22} \tilde{\rho}_{21} \\ \rho_{11} \tilde{\rho}_{12} + \rho_{12} \tilde{\rho}_{11} \end{pmatrix} +
g_U
\begin{pmatrix}
1 \\ 1
\end{pmatrix} v_{\phi_U}^2 = 0\,, \nonumber\\
&&(\rho\tilde{\rho})_{(\mathbf{1}'_L,\mathbf{1}_R)} = \rho_{11} \tilde{\rho}_{22} - \rho_{21} \tilde{\rho}_{12} + \rho_{12} \tilde{\rho}_{21} - \rho_{22} \tilde{\rho}_{11} = 0\,,
\end{eqnarray}
where the VEV of $\phi_U$, $\langle \phi_U \rangle=(0,1,-1)^T v_{\phi_U}$, has been used. These equations determine Eq.~\eqref{eq:VEV_rho} 
(or in turn) with 
\begin{eqnarray} \label{eq:chi1}
v_\rho v_{\tilde{\rho}} = - g_U v_{\phi_U}^2
\end{eqnarray} 
satisfied. 
Note that $v_\rho$ and $v_{\tilde{\rho}}$ cannot be determined by the above equation. They will be determined later once other minimisation conditions are satisfied. 

To achieve step II, the $U$-invariant direction of $\langle \rho \rangle$ and $\langle \tilde{\rho} \rangle$ in $S_{4L}$ is important. We write out driving terms to determine directions of $\langle \phi_{\rm sol} \rangle$ and $\langle \phi_{\rm atm} \rangle$ with $\rho$ and $\tilde{\rho}$ involved,
\begin{eqnarray} \label{eq:superpotential_driving7}
w_d \supset \phi^d_N \left[ (\phi_N \rho)_{(\mathbf{3}_L, \mathbf{1}'_R)} + g_t (\tilde{\phi}'_t \phi_U)_{(\mathbf{3}_L, \mathbf{1}'_R)} \right] + \phi^{d\prime}_N \left[ (\phi_N \tilde{\rho})_{(\mathbf{3}_L, \mathbf{1}_R)} + g_\eta \eta \phi_U \right] \,.
\end{eqnarray} 
Once $\rho$, $\tilde{\rho}$ and $\phi_U$ get the VEVs, the minimisation leads to 
\begin{eqnarray}
&&(\phi_N \rho)_{(\mathbf{3}_L, \mathbf{1}'_R)}|_{\langle \rho \rangle} + g_t (\tilde{\phi}'_t \phi_U)_{(\mathbf{3}_L, \mathbf{1}'_R)}|_{\langle \tilde{\phi}'_t \rangle,\langle \phi_U \rangle} = 
\begin{pmatrix} \phi_{\rm sol,2} + \phi_{\rm sol,3} \\ \phi_{\rm sol,3} + \phi_{\rm sol,1} \\ \phi_{\rm sol,1} + \phi_{\rm sol,2} \end{pmatrix} v_\rho - g_t \begin{pmatrix} 1 \\ 2 \\ 0 \end{pmatrix} v_{\tilde{\phi}'_t} v_{\phi_U} = 0 \,,\nonumber\\
&&(\phi_N \tilde{\rho})_{(\mathbf{3}_L, \mathbf{1}_R)}|_{\langle \tilde{\rho} \rangle} + g_\eta \eta \phi_U|_{\langle \phi_U \rangle,\langle \eta \rangle} =  
\begin{pmatrix} \phi_{\rm atm,2} + \phi_{\rm atm,3} \\ \phi_{\rm atm,3} + \phi_{\rm atm,1} \\ \phi_{\rm atm,1} + \phi_{\rm atm,2} \end{pmatrix} v_{\tilde{\rho}} + g_\eta \begin{pmatrix} 0 \\ 1 \\ -1 \end{pmatrix} v_\eta v_{\phi_U} = 0\,,
\end{eqnarray}
where $\phi_{\rm sol} = (\phi_{\rm sol,1}, \phi_{\rm sol,2},\phi_{\rm sol,3})^T$ and $\phi_{\rm atm} = (\phi_{\rm atm,1}, \phi_{\rm atm,2},\phi_{\rm atm,3})^T$. 
As shown above, the contractions $(\phi_N \rho)_{(\mathbf{3}_L, \mathbf{1}'_R)}$ and $(\phi_N \tilde{\rho})_{(\mathbf{3}_L, \mathbf{1}'_R)}$ select $\phi_{\rm sol}$ and $\phi_{\rm atm}$, respectively. After $\rho$ and $\tilde{\rho}$ gain their VEVs, $\phi_{\rm sol}$ and $\phi_{\rm atm}$ are separated and gain the required VEV directions in Eq.~\eqref{eq:VEV12} separately, with correlations
\begin{eqnarray}\label{eq:chi2}
2\frac{v_{\phi_{\rm sol}}}{\sqrt{11}} v_\rho = g_t v_{\tilde{\phi}'_t} v_{\phi_U} \,, \quad
\frac{v_{\phi_{\rm atm}}}{\sqrt{2}} v_{\tilde{\rho}} = - g_\eta v_\eta v_{\phi_U}
\end{eqnarray}
satisfied. Here, we have applied the technique developed in Ref.~\cite{King:2016yvg} to achieve the direction $(1,-1,3)^T$.

Finally, we consider how to achieve  $v_{\phi_{\rm sol}} = v_{\phi_{\rm atm}}$ in step III. We introduce another flavon $\tilde{\phi}_N \equiv (\tilde{\phi}_{\rm atm}, \tilde{\phi}_{\rm sol})$, which transforms as $(\mathbf{3}_L, \mathbf{2}_R)$, the same as $\phi_N$. Given the following driving terms similar to those for $\tilde{\phi}_N$, 
\begin{eqnarray} \label{eq:superpotential_driving8}
w_d \supset \tilde{\phi}^d_N \left[ (\tilde{\phi}_N \tilde{\rho})_{(\mathbf{3}_L, \mathbf{1}'_R)} + g_{\tilde{t}} (\phi'_t \phi_U)_{(\mathbf{3}_L, \mathbf{1}'_R)} \right] + \tilde{\phi}^{d\prime}_N \left[ (\tilde{\phi}_N \rho)_{(\mathbf{3}_L, \mathbf{1}'_R)} + g_{\tilde{\eta}} \tilde{\eta} \phi_U \right]
\end{eqnarray} 
and following a similar analysis, we arrive at
\begin{eqnarray} 
\langle \tilde{\phi}_{\rm atm}\rangle = \begin{pmatrix} 0 \\ \frac{1}{\sqrt{2}} \\ \frac{-1}{\sqrt{2}} \end{pmatrix} v_{\tilde{\phi}_{\rm atm}} \,, \quad
\langle \tilde{\phi}_{\rm sol} \rangle = \begin{pmatrix} \frac{1}{\sqrt{11}} \\ \frac{3}{\sqrt{11}} \\ \frac{-1}{\sqrt{11}} \end{pmatrix} v_{\tilde{\phi}_{\rm sol}} 
\end{eqnarray}
with $v_{\tilde{\phi}_{\rm sol}}$ and $v_{\tilde{\phi}_{\rm atm}}$ satisfying 
\begin{eqnarray}\label{eq:chi2}
2\frac{v_{\tilde{\phi}_{\rm sol}}}{\sqrt{11}} v_{\tilde{\rho}} = g_{\tilde{t}} v_{\phi'_t} v_{\phi_U} \,, \quad
\frac{v_{\tilde{\phi}_{\rm atm}}}{\sqrt{2}} v_{\rho} = - g_{\tilde{\eta}} v_{\tilde{\eta}} v_{\phi_U} \,.
\end{eqnarray}
Then, we construct the driving terms
\begin{eqnarray} \label{eq:superpotential_driving9}
w_d \supset \varphi^d_N (\phi_N \tilde{\phi}_N)_{(\mathbf{1}_L, \mathbf{1}'_R)} + \varphi^{d\prime}_N \left[ (\phi_N \tilde{\phi}_N)_{(\mathbf{1}_L, \mathbf{2}_R)} + A_\sigma \sigma \right] \,.
\end{eqnarray}
These terms result in
\begin{eqnarray}
&&(\phi_N \tilde{\phi}_N)_{(\mathbf{1}_L, \mathbf{1}'_R)} = v_{\phi_{\rm atm}} v_{\tilde{\phi}_{\rm atm}} - v_{\phi_{\rm sol}} v_{\tilde{\phi}_{\rm sol}} = 0 \,,\nonumber\\
&&(\phi_N \tilde{\phi}_N)_{(\mathbf{1}_L, \mathbf{2}_R)} + A_\sigma \sigma|_{\langle \sigma \rangle} = \frac{4}{\sqrt{22}}(v_{\phi_{\rm sol}} v_{\tilde{\phi}_{\rm atm}}, v_{\phi_{\rm atm}} v_{\tilde{\phi}_{\rm sol}}) = (1,1) A_\sigma v_\sigma\,,
\end{eqnarray}
where $\langle \sigma \rangle = (1,1)v_\sigma$ has been used. 
Following a straightforward calculation, we obtain 
\begin{eqnarray}
v_{\phi_{\rm atm}} = v_{\phi_{\rm sol}}\,,\quad v_{\tilde{\phi}_{\rm atm}} = v_{\tilde{\phi}_{\rm sol}}\,,\quad
v_{\phi_{\rm atm}} v_{\tilde{\phi}_{\rm atm}} = \frac{\sqrt{22}}{4} A_\sigma v_\sigma \,.
\end{eqnarray}
Combining the above equation with Eqs.~\eqref{eq:chi1} and \eqref{eq:chi2}, we further determine $v_\rho$ and $v_{\tilde{\rho}}$, 
\begin{eqnarray}
v_{\rho} = \sqrt{ \frac{11 g_t g_U v_{\tilde{\phi}'_t}}{2\sqrt{2} g_\eta v_\eta } } v_{\phi_U} \,,\quad 
v_{\tilde{\rho}} = \sqrt{  \frac{11 g_{\tilde{t}} g_U v_{\phi'_t}}{2\sqrt{2} g_{\tilde{\eta}} v_{\tilde{\eta}} } } v_{\phi_U} \,.
\end{eqnarray}


\subsection{Charge assignment of the model \label{sec:charge}}


\begin{table}[h!]
\centering
\begin{tabular}{cccccc|cccccc}\hline\hline
\multicolumn{2}{c}{Fields} & $S_{4 L}$ & $S_{4 R}$ & $U(1)$ & $U(1)'$  & 
 \multicolumn{2}{c}{Fields} & $S_{4 L}$ & $S_{4 R}$ & $U(1)$ & $U(1)'$  \\\hline
 \multirow{7}{*}{\rotatebox{90}{Higgs \& leptons}} & $\ell$  &  $\mathbf{3}_L$  &  $\mathbf{1}_R$  &  $1$  &  $6$ &
 \multirow{22}{*}{\rotatebox{90}{Driving \; fields}} & $\phi _T^d$  &  $\mathbf{2}_L$  &  $\mathbf{1}_R$  &  $-4$  &  $-4$ \\
& $e^c$  &  $\mathbf{1}'_L$  &  $\mathbf{1}'_R$  &  $-4$  &  $3$ &
 &  $\phi _S^d$  &  $\mathbf{3}_L$  &  $\mathbf{1}_R$  &  $4$  &  $-2$ \\
& $\mu ^c$  &  $\mathbf{1}_L$  &  $\mathbf{1}_R$  &  $-3$  &  $0$ &
 &  $\phi _U^d$  &  $\mathbf{3}'_L$  &  $\mathbf{1}_R$  &  $0$  &  $-3$ \\
& $\tau ^c$  &  $\mathbf{1}'_L$  &  $\mathbf{1}'_R$  &  $-2$  &  $-3$ &
 &  $\phi _t^d$  &  $\mathbf{1}_L$  &  $\mathbf{1}_R$  &  $-2$  &  $6$ \\
& $N^c$  &  $\mathbf{1}_L$  &  $\mathbf{2}_R$  &  $-4$  &  $-3$ &
 &  $\phi _t^{d\prime}$  &  $\mathbf{1}'_L$  &  $\mathbf{1}_R$  &  $-3$  &  $1$ \\
& $H_u$  &  $\mathbf{1}_L$  &  $\mathbf{1}_R$  &  $0$  &  $0$ &
 &  $\tilde{\phi }_t^d$  &  $\mathbf{1}_L$  &  $\mathbf{1}_R$  &  $-3$  &  $1$ \\
& $H_d$  &  $\mathbf{1}_L$  &  $\mathbf{1}_R$  &  $0$  &  $0$ &
 &  $\tilde{\phi }_t^{d\prime}$  &  $\mathbf{1}'_L$  &  $\mathbf{1}_R$  &  $-9$  &  $3$ \\
\cline{1-6}
 \multirow{15}{*}{\rotatebox{90}{Flavons}} &  $\phi _T$  &  $\mathbf{3}_L$  &  $\mathbf{1}_R$  &  $2$  &  $2$ &
 &  $\varphi _T^d$  &  $\mathbf{1}_L$  &  $\mathbf{1}_R$  &  $-8$  &  $8$ \\
& $\phi _S$  &  $\mathbf{3}_L$  &  $\mathbf{1}_R$  &  $-2$  &  $1$ &
 &  $\xi _T^d$  &  $\mathbf{1}_L$  &  $\mathbf{2}_R$  &  $12$  &  $-8$ \\
& $\phi _U$  &  $\mathbf{3}'_L$  &  $\mathbf{1}_R$  &  $0$  &  $3$ &
 &  $\xi _{TS}^d$  &  $\mathbf{1}_L$  &  $\mathbf{2}_R$  &  $-10$  &  $-6$ \\
& $\phi _t^{\prime}$  &  $\mathbf{3}'_L$  &  $\mathbf{1}'_R$  &  $1$  &  $-3$ &
 &  $\xi _t^d$  &  $\mathbf{1}_L$  &  $\mathbf{1}_R$  &  $-6$  &  $-6$ \\
& $\tilde{\phi }_t^{\prime}$  &  $\mathbf{3}'_L$  &  $\mathbf{1}'_R$  &  $7$  &  $-5$ &
 &  $\xi _t^{d\prime}$  &  $\mathbf{1}_L$  &  $\mathbf{1}'_R$  &  $3$  &  $-7$ \\
& $\xi _T$  &  $\mathbf{1}_L$  &  $\mathbf{3}_R$  &  $-6$  &  $4$ &
 &  $\rho ^d$  &  $\mathbf{2}_L$  &  $\mathbf{1}_R$  &  $-8$  &  $-2$ \\
& $\xi _{TS}$  &  $\mathbf{1}_L$  &  $\mathbf{3}_R$  &  $5$  &  $3$ &
 &  $\tilde{\rho }^d$  &  $\mathbf{2}_L$  &  $\mathbf{1}_R$  &  $8$  &  $-10$ \\
& $\xi _t^{\prime}$  &  $\mathbf{1}_L$  &  $\mathbf{3}'_R$  &  $3$  &  $3$ &
 &  $\rho ^{d\prime}$  &  $\mathbf{2}_L$  &  $\mathbf{1}_R$  &  $0$  &  $-6$ \\
& $\rho $  &  $\mathbf{2}_L$  &  $\mathbf{2}_R$  &  $4$  &  $1$ &
 &  $\rho ^{d\prime\prime}$  &  $\mathbf{1}'_L$  &  $\mathbf{1}_R$  &  $0$  &  $-6$ \\
& $\tilde{\rho }$  &  $\mathbf{2}_L$  &  $\mathbf{2}_R$  &  $-4$  &  $5$ &
 &  $\phi _N^d$  &  $\mathbf{3}_L$  &  $\mathbf{1}'_R$  &  $-7$  &  $2$ \\
& $\phi _N$  &  $\mathbf{3}_L$  &  $\mathbf{2}_R$  &  $3$  &  $-3$ &
 &  $\phi _N^{d\prime}$  &  $\mathbf{3}_L$  &  $\mathbf{1}'_R$  &  $1$  &  $-2$ \\
& $\tilde{\phi }_N$  &  $\mathbf{3}_L$  &  $\mathbf{2}_R$  &  $5$  &  $-5$ &
 &  $\tilde{\phi }_N^d$  &  $\mathbf{3}_L$  &  $\mathbf{1}'_R$  &  $-1$  &  $0$ \\
& $\eta $  &  $\mathbf{1}'_L$  &  $\mathbf{1}_R$  &  $-1$  &  $-1$ &
 &  $\tilde{\phi }_N^{d\prime}$  &  $\mathbf{3}_L$  &  $\mathbf{1}'_R$  &  $-9$  &  $4$ \\
& $\tilde{\eta }$  &  $\mathbf{1}'_L$  &  $\mathbf{1}_R$  &  $9$  &  $-7$ &
 &  $\varphi _N^d$  &  $\mathbf{1}_L$  &  $\mathbf{1}'_R$  &  $-8$  &  $8$ \\
& $\sigma $  &  $\mathbf{1}_L$  &  $\mathbf{2}_R$  &  $8$  &  $-8$ &
 &  $\varphi _N^{d\prime}$  &  $\mathbf{1}_L$  &  $\mathbf{2}_R$  &  $-8$  &  $8$ \\\hline\hline
\end{tabular} 
  \caption{\label{tab:fields} Field arrangements of the $\mu\tau$-LSS model in $S_{4L} \times S_{4R} \times U(1) \times U(1)'$. In addition, we assume a standard $U(1)_R$ symmetry with the charge assignments: +1 for lepton, 0 for Higgs and flavon fields, and +2 for driving fields. }
\end{table}

Finally, we list our particle content in Table~\ref{tab:fields}. Representations of all fields in $S_{4L} \times S_{4R}$ are explicitly the same as introduced in the last subsection. In order to forbid unnecessary terms, e.g., $\phi_t^d \phi_U \phi_U$, which may violate the required directions but cannot be forbidden by two $S_4$'s, we introduce two $U(1)$ symmetries. The unnecessary Goldstone bosons accompanying with $U(1)$ breakings can be avoided by considering small $U(1)$-explicit-breaking terms, which will not be discussed here. Table~\ref{tab:fields} guarantees superpotential terms $w_\ell$ and $w_\nu$ in Eq.~\eqref{eq:superpotential_leptons} to generate lepton masses. No extra higher-dimensional operators up to $d=6$ should be considered. These $U(1)$ symmetries can in principle be replaced by several $Z_n$ symmetries with a careful arrangement of all field charges. For the driving superpotential, we only consider renormalisable terms. The full driving superpotential $w_d$ is the collection of Eqs.~\eqref{eq:superpotential_driving1}, \eqref{eq:superpotential_driving2}, \eqref{eq:superpotential_driving3}, \eqref{eq:superpotential_driving4}, \eqref{eq:superpotential_driving5}, \eqref{eq:superpotential_driving6}, \eqref{eq:superpotential_driving7}, \eqref{eq:superpotential_driving8}, and \eqref{eq:superpotential_driving9}, but re-expressing $\mu_x^2$ in Eq.~\eqref{eq:superpotential_driving4} as
\begin{eqnarray}
\mu_x^2 = g_{t1} \eta \tilde{\eta} + g_{t2} (\phi_N \phi_N)_{(\mathbf{1}_L,\mathbf{1}_R)} \,,
\end{eqnarray} 
where $g_{t1}$ and $g_{t2}$ are dimensionless coefficients.

\section{Conclusion \label{sec:conclusion}}

In this paper we have proposed a $\mu-\tau$ reflection symmetric Littlest Seesaw ($\mu\tau$-LSS) model.
In this model the two mass parameters of the LSS model are fixed to be in a special 
ratio by symmetry, so that the resulting neutrino mass matrix in the flavour basis
(after the seesaw mechanism has been applied) satisfies
$\mu-\tau$ reflection symmetry and 
has only one free adjustable parameter, namely an overall free mass scale. 

The resulting $\mu\tau$-LSS model predicts  $\theta_{23} =45^\circ$, $\delta=-90^\circ$ and $\theta_{12} \approx 34.5^\circ$, which are compatible with data. The predicted $\theta_{13}$ and the  ratio of mass square differences $\Delta m^2_{21}/\Delta m^2_{31}$ are out of the $3\sigma$ ranges of the current global oscillation data. 
However, with radiative corrections included, assuming SUSY,
all mixing parameters and the ratio $\Delta m^2_{21}/\Delta m^2_{31}$ depend on one single free parameter, namely $\epsilon$, which can bring all the observables within their $3\sigma$ ranges.

We have constructed a concrete lepton flavour model in $S_{4L} \times S_{4R} \times U(1) \times U(1)'$ to realise littlest mu-tau seesaw model, $S_{4L}$ for left-handed fermions and $S_{4R}$ for right-handed fermions.
The two right-handed neutrinos are arranged as singlets in $S_{4L}$, in usual constrained sequential dominance.
However they are arranged as a doublet $S_{4R}$, which is necessary to achieve the desired ratio of effective
mass parameters $\frac{m_a}{m_s}=11$ as required for $\mu-\tau$ reflection symmetry.

The desired ratio of effective
mass parameters $\frac{m_a}{m_s}=11$ also relies on special 
vacuum alignments which have been carefully realised with the help of SUSY
driving fields. Specifically, the flavon $\phi_N = (\phi_{\rm atm}, \phi_{\rm sol})$, which contributing to the neutrino Dirac mass matrix, achieves VEV $\langle \phi_{\rm sol} \rangle = (1,-1,3)^T v_{\phi_{\rm sol}}/\sqrt{11}$ and $\langle \phi_{\rm atm} \rangle = (0,-1,1)^T v_{\phi_{\rm atm}}/\sqrt{2}$ separately in $S_{4L}$ and $v_{\phi_{\rm sol}} = v_{\phi_{\rm atm}}$ due to the constraint of $S_{4R}$. 

Although the high energy model is rather complicated, involving $(S_4\times U(1))^2$,
with many flavons and driving fields, the low energy neutrino mass matrix 
has ultimate simplicity, with a built-in $\mu-\tau$ reflection symmetry and tri-maximal mixing.
Since the neutrino mass matrix only depends on one overall mass scale,
the low energy observables are completely specified in terms of one radiative correction parameter,
leading to testable predictions for all lepton mixing angles and CP phases,
as well as neutrino mass ratios.

\subsection*{Acknowledgements}

We acknowledge the STFC Consolidated Grant ST/L000296/1 and the European Union's Horizon 2020 Research and Innovation programme under Marie Sk\l{}odowska-Curie grant agreements Elusives ITN No.\ 674896 and InvisiblesPlus RISE No.\ 690575. Y.\,L.\,Z. thanks Z.\,Z.\,Xing for sharing RG running code.


\appendix
\section{Group theory of $S_4$ \label{app:S4}} 

$S_4$ is the permutation group of 4 objects. Its three generators $S$, $T$ and $U$ satisfying the equalities $T^3=S^2=U^2=(ST)^3=(SU)^2=(TU)^2=1$, from which $(STU)^4=1$ is automatically obtained. The minimal number of generators of $S_4$ is actually two \cite{Bazzocchi:2009pv, Ding:2013eca}. However, we follow the presentation in \cite{deMedeirosVarzielas:2017hen} to emphasise the $Z_2$ residual symmetries generated by $S$ and $U$, respectively. $S_4$ contains 5 irreducible representations (irreps), $\mathbf{1}$, $\mathbf{1}'$, $\mathbf{2}$, $\mathbf{3}$ and $\mathbf{3}'$. Throughout this paper, we work in the basis where the generator $T$ diagonal. Generators of $S_4$ in different irreps are listed in Table \ref{tab:rep_matrix2}.  

\begin{table}[h!]
\begin{center}
\begin{tabular}{cccc}
\hline\hline
   & $T$ & $S$ & $U$  \\\hline
$\mathbf{1}$ & 1 & 1 & 1 \\
$\mathbf{1^{\prime}}$ & 1 & 1 & $-1$ \\
$\mathbf{2}$ & 
$\left(
\begin{array}{cc}
 \omega  & 0 \\
 0 & \omega ^2 \\
\end{array}
\right)$ & 
$\left(
\begin{array}{cc}
 1 & 0 \\
 0 & 1 \\
\end{array}
\right)$ & 
$\left(
\begin{array}{cc}
 0 & 1 \\
 1 & 0 \\
\end{array}
\right)$ \\

$\mathbf{3}$ &  $\left(
\begin{array}{ccc}
 1 & 0 & 0 \\
 0 & \omega ^2 & 0 \\
 0 & 0 & \omega  \\
\end{array}
\right)$ &
$\frac{1}{3} \left(
\begin{array}{ccc}
 -1 & 2 & 2 \\
 2 & -1 & 2 \\
 2 & 2 & -1 \\
\end{array}
\right)$ &
$\left(
\begin{array}{ccc}
 1 & 0 & 0 \\
 0 & 0 & 1 \\
 0 & 1 & 0 \\
\end{array}
\right)$ \\

$\mathbf{3^{\prime}}$ &  $\left(
\begin{array}{ccc}
 1 & 0 & 0 \\
 0 & \omega ^2 & 0 \\
 0 & 0 & \omega  \\
\end{array}
\right)$ &
$\frac{1}{3} \left(
\begin{array}{ccc}
 -1 & 2 & 2 \\
 2 & -1 & 2 \\
 2 & 2 & -1 \\
\end{array}
\right)$ &
$-\left(
\begin{array}{ccc}
 1 & 0 & 0 \\
 0 & 0 & 1 \\
 0 & 1 & 0 \\
\end{array}
\right)$ \\ \hline\hline

\end{tabular}
\caption{\label{tab:rep_matrix2} The representation matrices for the $S_4$ generators $T$, $S$ and $U$, where $\omega=e^{2\pi i/3}$.}
\end{center}
\end{table}

The Kronecker products between different irreps can be easily obtained:
\begin{eqnarray}
&&\hspace{-1cm}
\mathbf{1}\otimes\mathbf{r}=\mathbf{r},~
\mathbf{1^{\prime}}\otimes\mathbf{1^{\prime}}=\mathbf{1}, ~
\mathbf{1^{\prime}}\otimes\mathbf{2}=\mathbf{2}, ~
\mathbf{1^{\prime}}\otimes\mathbf{3}=\mathbf{3^{\prime}}, ~
\mathbf{1^{\prime}}\otimes\mathbf{3^{\prime}}=\mathbf{3},~
\mathbf{2}\otimes\mathbf{2}=\mathbf{1}\oplus\mathbf{1}^{\prime}\oplus\mathbf{2},\nonumber\\
&&\hspace{-1cm}
\mathbf{2}\otimes\mathbf{3}=\mathbf{2}\otimes\mathbf{3^{\prime}}=\mathbf{3}\oplus\mathbf{3}^{\prime},\,
\mathbf{3}\otimes\mathbf{3}=\mathbf{3^{\prime}}\otimes\mathbf{3^{\prime}}=\mathbf{1}\oplus \mathbf{2}\oplus\mathbf{3}\oplus\mathbf{3^{\prime}},\,
\mathbf{3}\otimes\mathbf{3^{\prime}}=\mathbf{1^{\prime}}\oplus \mathbf{2}\oplus\mathbf{3}\oplus\mathbf{3^{\prime}},
\end{eqnarray}
where $\mathbf{r}$ represents any irrep of $S_4$. 
The Clebsch-Gordan coefficients for products of any two irreps $a$ and $b$ are listed in Table~\ref{tab:S4_CG}.

\begin{table}[h!]
\begin{center}
\begin{tabular}{lll}
\hline\hline
$\bullet$ & $a\sim\mathbf{1}, b\sim \mathbf{r}$ &
$(ab)_{\mathbf{r}}= a (b_1, b_2, ...)$ \\[2mm]
$\bullet$ & $a\sim b \sim \mathbf{1}'$ &
$(ab)_{\mathbf{1}} = a b$ \\[2mm]
$\bullet$ & $a\sim \mathbf{1}', b \sim \mathbf{2}$ &
$(ab)_{\mathbf{2}} = a (b_1, -b_2)$ \\[2mm]
$\bullet$ & $a\sim \mathbf{1}', b \sim \mathbf{3} (\mathbf{3'})$ &
$(ab)_{\mathbf{3}'(\mathbf{3})} = a (b_1, b_2, b_3)$ \\[2mm]
$\bullet$ & $a \sim b \sim\mathbf{2}$ 
& $(ab)_\mathbf{1} = a_1b_2 + a_2b_1$ \\
&& $(ab)_\mathbf{1'} = a_1b_2 - a_2b_1$ \\
&& $(ab)_{\mathbf{2}} = (a_2b_2, a_1b_1)$  \\[2mm]
$\bullet$ & $a\sim\mathbf{2}, b\sim\mathbf{3} (\mathbf{3}')$ 
& $(ab)_{\mathbf{3}  (\mathbf{3}')} = (a_1b_2+a_2b_3, a_1b_3+a_2b_1, a_1b_1+a_2b_2)$ \\
&& $(ab)_{\mathbf{3}'  (\mathbf{3})} = (a_1b_2-a_2b_3, a_1b_3-a_2b_1, a_1b_1-a_2b_2)$ \\[2mm]
$\bullet$ & $a \sim b\sim\mathbf{3} \text{ or } \mathbf{3}'$ 
& $(ab)_\mathbf{1} = a_1b_1 + a_2b_3 + a_3b_2$ \\
&& $(ab)_\mathbf{2} = (a_2b_2 + a_1b_3 + a_3b_1,~ a_3b_3 + a_1b_2 + a_2b_1)$ \\
&& $(ab)_{\mathbf{3}} = (2a_1b_1\!-\!a_2b_3\!-\!a_3b_2, 2a_3b_3\!-\!a_1b_2\!-\!a_2b_1, 2a_2b_2\!-\!a_3b_1\!-\!a_1b_3)$ \\
&& $(ab)_{\mathbf{3}'} = (a_2b_3-a_3b_2, a_1b_2-a_2b_1, a_3b_1-a_1b_3)$ \\[2mm]
$\bullet$ & $a \sim \mathbf{3}$,  $b\sim \mathbf{3}'$ 
& $(ab)_{\mathbf{1}'} = a_1b_1 + a_2b_3 + a_3b_2$ \\
&& $(ab)_\mathbf{2} = (a_2b_2 + a_1b_3 + a_3b_1,~ -a_3b_3 - a_1b_2 - a_2b_1)$ \\
&& $(ab)_{\mathbf{3}} = (a_2b_3-a_3b_2, a_1b_2-a_2b_1, a_3b_1-a_1b_3)$\\
&& $(ab)_{\mathbf{3}'} = (2a_1b_1\!-\!a_2b_3\!-\!a_3b_2, 2a_3b_3\!-\!a_1b_2\!-\!a_2b_1, 2a_2b_2\!-\!a_3b_1\!-\!a_1b_3)$ \\\hline\hline
\end{tabular}
\caption{Kronecker products and Clebsch-Gordan (CG) coefficients of $S_4$. $\mathbf{r}$ represents any irrep of $S_4$. } \label{tab:S4_CG}
\end{center}
\end{table}

\section{Vacuum degeneracy \label{app:degeneracy}}

Degenerate vacua exist in theories of discrete symmetries. Discussion on structures and physical equivalence of degenerate vacua has been given in e.g., \cite{Morozumi:2017rrg} based on the $A_4$ symmetry. 
This appendix is devoted to the discussion of vacuum degeneracy in $S_4$. 
In section~\ref{sec:vacuum}, we select the VEV for $\phi_T$ along $(1,0,0)^T$ direction in $S_{4L}$, and those for $\xi_T$ and $\xi_{TS}$ along the $(1,0,0)$ and $(\frac{1}{3},-\frac{2}{3}\omega,-\frac{2}{3}\omega^2)$ directions in $S_{4R}$, respectively. Their validity is explained in the following.
Given any $S_4$ triplet flavon $\Phi = (\Phi_1, \Phi_2, \Phi_3)^T$ and the superpotential $w=\Phi^d (\Phi \Phi)_{\mathbf{2}}$ with $\Phi^d$ being an $S_4$ doublet driving field. The vacuum is solved via $\partial w / \partial \Phi^d = (\Phi \Phi)_{\mathbf{2}} = 0$, i.e.,
\begin{eqnarray}
\frac{\partial w}{\partial \Phi^d_1} &=& \Phi_2^2 + 2 \Phi_1 \Phi_3 = 0 \nonumber\\
\frac{\partial w}{\partial \Phi^d_2} &=& \Phi_3^2 + 2 \Phi_1 \Phi_2 = 0
\end{eqnarray} 
Straightforward calculation shows the full solution is given by
\begin{eqnarray}\label{eq:full_VEVs}
\hspace{-5mm}
\langle\Phi\rangle^{T} = \begin{pmatrix} 1 \\ 0 \\ 0 \end{pmatrix} v_{\Phi}, \,
\langle\Phi\rangle^{STS} = \begin{pmatrix} \frac{1}{3} \\ -\frac{2}{3} \\ -\frac{2}{3} \end{pmatrix} v_{\Phi},\,
\langle\Phi\rangle^{TS} = \begin{pmatrix} \frac{1}{3} \\ -\frac{2}{3}\omega \\ -\frac{2}{3}\omega^2 \end{pmatrix} v_{\Phi},\,
\langle\Phi\rangle^{ST} = \begin{pmatrix} \frac{1}{3} \\ -\frac{2}{3}\omega^2 \\ -\frac{2}{3}\omega \end{pmatrix} v_{\Phi}
\end{eqnarray}
with $v_{\Phi}$ undetermined. These VEVs are invariant under the transformation of $T$, $STS$, $TS$ and $ST$, respectively, 
\begin{eqnarray} \label{eq:VEV_invariant}
T\; \langle\Phi\rangle^T = \langle\Phi\rangle^T\,, \;
STS\; \langle\Phi\rangle^{STS} = \langle\Phi\rangle^{STS}\,, \;
TS\; \langle\Phi\rangle^{TS} = \langle\Phi\rangle^{TS}\,, \;
ST\; \langle\Phi\rangle^{ST} = \langle\Phi\rangle^{ST}\,,
\end{eqnarray}
and therefore, preserve residual symmetries generated by these elements,
\begin{eqnarray}
Z_3^T \!= \{ 1, T, T^2 \},
Z_3^{STS} \!= \{ 1, STS, ST^2S \},
Z_3^{TS} \!= \{ 1, TS, (TS)^2 \},
Z_3^{ST} \!= \{ 1, ST, (ST)^2 \}, \nonumber\\
\end{eqnarray}
respectively. Note that all these $Z_3$ symmetries are conjugate with each other. Their elements satisfy the following conjugacy transformations
\begin{eqnarray}
STS = S \; T \; S^{-1}\,,~
TS = (ST)^{-1} \; T \; (ST) \,,~
ST = (TS) \; T \; (TS)^{-1} \,.
\end{eqnarray}
Starting from one VEV, e.g., $\langle\Phi\rangle^T$, the rest degenerate VEVs are obtained via
\begin{eqnarray}
T\langle\Phi\rangle^T = \langle\Phi\rangle^T \Rightarrow \left\{ \begin{array}{ccc} 
STS\;S\langle\Phi\rangle^T = S\langle\Phi\rangle^T & \Rightarrow & \langle\Phi\rangle^{STS} = S\langle\Phi\rangle^T \\
(ST)^{-1} T (ST)\;(ST)^{-1}\langle\Phi\rangle^T = (ST)^{-1}\langle\Phi\rangle^T & \Rightarrow & \langle\Phi\rangle^{TS} = (ST)^{-1}\Phi^T \\
(TS) T (TS)^{-1}\;(ST)^{-1}\langle\Phi\rangle^T = (TS)\langle\Phi\rangle^T & \Rightarrow & \langle\Phi\rangle^{ST} = (TS)\langle\Phi\rangle^T
 \end{array} \right.\,.\nonumber\\
 \label{eq:conjugacy}
\end{eqnarray}
Therefore, one vacuum maps to another under the transformation of some group element.

In flavour model building, if there is only one flavon whose VEV preserves a $Z_3$, selecting one or another VEV, e.g., $\langle\Phi\rangle^T$ or $\langle\Phi\rangle^{STS}$, respectively, does not make a difference in the physical point of view. All VEVs connect with each other via the conjugacy transformation as discussed above. In other words, starting from $\langle\Phi\rangle^{STS}$, one can rotate it to $\langle\Phi\rangle^{T}$ via the reverse transformation of Eq.~\eqref{eq:conjugacy}. Therefore, it is always safe to use $\langle\Phi\rangle^T$ as the VEV for model building. This is what we have done for the VEV of $\phi_T$, which is the only $Z_3$-invariant VEV in $S_{4L}$. 

However, If there are two flavons $\Phi$ and $\Phi'$ taking $Z_3$-invariant VEVs at the same time, we must be careful about the VEV selection. Whether these VEVs preserve the same $Z_3$ or different $Z_3$'s are physically different. 

The number of degenerate VEVs is four, as shown in~\eqref{eq:full_VEVs}. From a naive estimation, we know that there is a propability of 1/4 that both $\langle\Phi\rangle$ and $\langle\Phi'\rangle$ preserve the same $Z_3$. In this case, one can always rotate them to the $Z_3^{T}$ one following the above discussion. 

For the other case, with a probability of 3/4, two flavon VEVs preserve different $Z_3$ symmetries. We argue that the discussion in the case that only one flavon VEV preserves a $Z_3$ does not hold. Instead, one can always rotate the $\Phi$ VEV to preserve $Z_3^T$ and the $\Phi'$ VEV to preserve $Z_3^{TS}$. We explain this with the help of the following example. Without loss of generality, let us assume the VEV of $\Phi$ preserves $Z_3^{STS}$ and that of $\Phi'$ preserves another one, $Z_3^{TS}$. Following the reverse transformation of Eq.~\eqref{eq:conjugacy}, the first flavon VEV $\langle\Phi\rangle^{STS}$ can always be rotated to $\langle\Phi\rangle^T$, 
\begin{eqnarray}
S^{-1}\,\langle\Phi\rangle^{STS} = \langle\Phi\rangle^T
\end{eqnarray}
The same transformation acting on $\langle\Phi'\rangle^{TS}$
\begin{eqnarray}
S^{-1}\,\langle\Phi'\rangle^{TS} = S^{-1} (ST)^{-1}\langle\Phi'\rangle^{T} = (TS)\,T\,\langle\Phi'\rangle^{T} = (TS)\, \langle\Phi'\rangle^{T} = \langle\Phi'\rangle^{ST}\,.
\end{eqnarray}
Here, for the first and the fourth identities, we have applied Eq.~\eqref{eq:conjugacy}, and for the third identity, we applied Eq.~\eqref{eq:VEV_invariant}. Once $\langle\Phi\rangle^T$ has been fixed, one can perform rotations by acting $T$ and $T^2$ which do not change the $Z_3^T$-invariant VEV $\langle\Phi\rangle^T$, but connect the $Z_3^{ST}$-invariant $\langle\Phi'\rangle^{ST}$ with $Z_3^{TS}$- and $Z_3^{STS}$-invariant VEVs, respectively, 
\begin{eqnarray}
T\, \langle\Phi'\rangle^{ST} = T (TS)\, \langle\Phi'\rangle^{T} = (ST)^{-1}\, \langle\Phi'\rangle^{T} = \langle\Phi'\rangle^{TS}\,, \nonumber\\
T^2\, \langle\Phi'\rangle^{ST} = T^2 (TS)\, \langle\Phi'\rangle^{T} = S\, \langle\Phi'\rangle^{T} = \langle\Phi'\rangle^{STS}\,. 
\end{eqnarray}
Therefore, under the transformation of $T$ and $T^2$, the rest three VEVs are physically equivalent. One can always select the $Z_3^{TS}$-invariant VEV for $\Phi'$. Applying this conclusion to our VEV alignment for $\xi_T$ and $\xi_{TS}$, we have both VEVs of $\xi_T$ and $\xi_{TS}$ preserving $Z_3$ symmetries of $S_{4R}$, a smaller chance that both of them select the $(1,0,0)$ direction and a larger chance that $\langle \xi_T \rangle$ selects $(1,0,0)$ direction and $\langle \xi_{TS} \rangle$ selects $(\frac{1}{3},-\frac{2}{3}\omega,-\frac{2}{3}\omega^2)$ direction. Involving more flavons may complicate the vacuum degeneracy problem and decrease the chance to achieve the required VEV, which will not be expanded here.


\begin{thebibliography}{99}

\bibitem{Xing:2011zza} 
  Z.~z.~Xing and S.~Zhou,
  Springer-Verlag, Berlin Heidelberg (2011)


\bibitem{King:2013eh} 
  S.~F.~King and C.~Luhn,
  Rept.\ Prog.\ Phys.\  {\bf 76}, 056201 (2013)
  doi:10.1088/0034-4885/76/5/056201
  [arXiv:1301.1340 [hep-ph]].


\bibitem{King:2015aea} 
  S.~F.~King,
  J.\ Phys.\ G {\bf 42}, 123001 (2015)
  doi:10.1088/0954-3899/42/12/123001
  [arXiv:1510.02091 [hep-ph]].


\bibitem{Capozzi:2018ubv} 
  F.~Capozzi, E.~Lisi, A.~Marrone and A.~Palazzo,
  Prog.\ Part.\ Nucl.\ Phys.\  {\bf 102}, 48 (2018)
  doi:10.1016/j.ppnp.2018.05.005
  [arXiv:1804.09678 [hep-ph]].


\bibitem{deSalas:2017kay} 
  P.~F.~de Salas, D.~V.~Forero, C.~A.~Ternes, M.~Tortola and J.~W.~F.~Valle,
  Phys.\ Lett.\ B {\bf 782}, 633 (2018)
  doi:10.1016/j.physletb.2018.06.019
  [arXiv:1708.01186 [hep-ph]].


\bibitem{Esteban:2016qun} 
  I.~Esteban, M.~C.~Gonzalez-Garcia, M.~Maltoni, I.~Martinez-Soler and T.~Schwetz,
  JHEP {\bf 1701}, 087 (2017)
  doi:10.1007/JHEP01(2017)087
  [arXiv:1611.01514 [hep-ph]].


\bibitem{Xing:2015fdg} 
  Z.~z.~Xing and Z.~h.~Zhao,
  Rept.\ Prog.\ Phys.\  {\bf 79}, no. 7, 076201 (2016)
  doi:10.1088/0034-4885/79/7/076201
  [arXiv:1512.04207 [hep-ph]].


\bibitem{Minkowski:1977sc} 
  P.~Minkowski,
  Phys.\ Lett.\  {\bf 67B}, 421 (1977).
  doi:10.1016/0370-2693(77)90435-X


\bibitem{Yanagida:1979as} 
  T.~Yanagida,
  Conf.\ Proc.\ C {\bf 7902131}, 95 (1979).


\bibitem{GellMann:1980vs} 
  M.~Gell-Mann, P.~Ramond and R.~Slansky,
  Conf.\ Proc.\ C {\bf 790927}, 315 (1979)
  [arXiv:1306.4669 [hep-th]].


\bibitem{Glashow:1979nm} 
  S.~L.~Glashow,
  NATO Sci.\ Ser.\ B {\bf 61}, 687 (1980).
  doi:10.1007/978-1-4684-7197-7\_15


\bibitem{Mohapatra:1979ia} 
  R.~N.~Mohapatra and G.~Senjanovic,
  Phys.\ Rev.\ Lett.\  {\bf 44}, 912 (1980).
  doi:10.1103/PhysRevLett.44.912


\bibitem{King:1999mb} 
  S.~F.~King,
  Nucl.\ Phys.\ B {\bf 576}, 85 (2000)
  doi:10.1016/S0550-3213(00)00109-7
  [hep-ph/9912492].


\bibitem{King:2002nf} 
  S.~F.~King,
  JHEP {\bf 0209}, 011 (2002)
  doi:10.1088/1126-6708/2002/09/011
  [hep-ph/0204360].


\bibitem{Frampton:2002qc} 
  P.~H.~Frampton, S.~L.~Glashow and T.~Yanagida,
  Phys.\ Lett.\ B {\bf 548}, 119 (2002)
  doi:10.1016/S0370-2693(02)02853-8
  [hep-ph/0208157].


\bibitem{Fukugita:1986hr} 
  M.~Fukugita and T.~Yanagida,
  Phys.\ Lett.\ B {\bf 174}, 45 (1986).
  doi:10.1016/0370-2693(86)91126-3


\bibitem{Guo:2003cc} 
  W.~l.~Guo and Z.~z.~Xing,
  Phys.\ Lett.\ B {\bf 583}, 163 (2004)
  doi:10.1016/j.physletb.2003.12.043
  [hep-ph/0310326].


\bibitem{Ibarra:2003up} 
  A.~Ibarra and G.~G.~Ross,
  Phys.\ Lett.\ B {\bf 591}, 285 (2004)
  doi:10.1016/j.physletb.2004.04.037
  [hep-ph/0312138].


\bibitem{Mei:2003gn} 
  J.~w.~Mei and Z.~z.~Xing,
  Phys.\ Rev.\ D {\bf 69}, 073003 (2004)
  doi:10.1103/PhysRevD.69.073003
  [hep-ph/0312167].


\bibitem{Guo:2006qa} 
  W.~l.~Guo, Z.~z.~Xing and S.~Zhou,
  Int.\ J.\ Mod.\ Phys.\ E {\bf 16}, 1 (2007)
  doi:10.1142/S0218301307004898
  [hep-ph/0612033].


\bibitem{Antusch:2011nz} 
  S.~Antusch, P.~Di Bari, D.~A.~Jones and S.~F.~King,
  Phys.\ Rev.\ D {\bf 86}, 023516 (2012)
  doi:10.1103/PhysRevD.86.023516
  [arXiv:1107.6002 [hep-ph]].


\bibitem{Harigaya:2012bw} 
  K.~Harigaya, M.~Ibe and T.~T.~Yanagida,
  Phys.\ Rev.\ D {\bf 86}, 013002 (2012)
  doi:10.1103/PhysRevD.86.013002
  [arXiv:1205.2198 [hep-ph]].


\bibitem{Zhang:2015tea} 
  J.~Zhang and S.~Zhou,
  JHEP {\bf 1509}, 065 (2015)
  doi:10.1007/JHEP09(2015)065
  [arXiv:1505.04858 [hep-ph]].


\bibitem{King:2013iva} 
  S.~F.~King,
  JHEP {\bf 1307}, 137 (2013)
  doi:10.1007/JHEP07(2013)137
  [arXiv:1304.6264 [hep-ph]].


\bibitem{Bjorkeroth:2014vha} 
  F.~Bj\"{o}rkeroth and S.~F.~King,
  J.\ Phys.\ G {\bf 42}, no. 12, 125002 (2015)
  doi:10.1088/0954-3899/42/12/125002
  [arXiv:1412.6996 [hep-ph]].


\bibitem{King:2015dvf} 
  S.~F.~King,
  JHEP {\bf 1602}, 085 (2016)
  doi:10.1007/JHEP02(2016)085
  [arXiv:1512.07531 [hep-ph]].


\bibitem{Bjorkeroth:2015ora} 
  F.~Bj\"{o}rkeroth, F.~J.~de Anda, I.~de Medeiros Varzielas and S.~F.~King,
  JHEP {\bf 1506}, 141 (2015)
  doi:10.1007/JHEP06(2015)141
  [arXiv:1503.03306 [hep-ph]].


\bibitem{Bjorkeroth:2015tsa} 
  F.~Bj\"{o}rkeroth, F.~J.~de Anda, I.~de Medeiros Varzielas and S.~F.~King,
  JHEP {\bf 1510}, 104 (2015)
  doi:10.1007/JHEP10(2015)104
  [arXiv:1505.05504 [hep-ph]].


\bibitem{King:2016yvg} 
  S.~F.~King and C.~Luhn,
  JHEP {\bf 1609}, 023 (2016)
  doi:10.1007/JHEP09(2016)023
  [arXiv:1607.05276 [hep-ph]].


\bibitem{Ballett:2016yod} 
  P.~Ballett, S.~F.~King, S.~Pascoli, N.~W.~Prouse and T.~Wang,
  JHEP {\bf 1703}, 110 (2017)
  doi:10.1007/JHEP03(2017)110
  [arXiv:1612.01999 [hep-ph]].


\bibitem{King:2018kka} 
  S.~F.~King and C.~C.~Nishi,
  Phys.\ Lett.\ B {\bf 785}, 391 (2018)
  doi:10.1016/j.physletb.2018.08.056
  [arXiv:1807.00023 [hep-ph]].

\bibitem{Xing:2006ms} 
  Z.~z.~Xing and S.~Zhou,
  Phys.\ Lett.\ B {\bf 653}, 278 (2007)
  doi:10.1016/j.physletb.2007.08.009
  [hep-ph/0607302].


\bibitem{Albright:2008rp} 
  C.~H.~Albright and W.~Rodejohann,
  Eur.\ Phys.\ J.\ C {\bf 62}, 599 (2009)
  doi:10.1140/epjc/s10052-009-1074-3
  [arXiv:0812.0436 [hep-ph]].


\bibitem{Albright:2010ap} 
  C.~H.~Albright, A.~Dueck and W.~Rodejohann,
  Eur.\ Phys.\ J.\ C {\bf 70}, 1099 (2010)
  doi:10.1140/epjc/s10052-010-1492-2
  [arXiv:1004.2798 [hep-ph]].


\bibitem{He:2011gb} 
  X.~G.~He and A.~Zee,
  Phys.\ Rev.\ D {\bf 84}, 053004 (2011)
  doi:10.1103/PhysRevD.84.053004
  [arXiv:1106.4359 [hep-ph]].


\bibitem{Rodejohann:2012cf} 
  W.~Rodejohann and H.~Zhang,
  Phys.\ Rev.\ D {\bf 86}, 093008 (2012)
  doi:10.1103/PhysRevD.86.093008
  [arXiv:1207.1225 [hep-ph]].


\bibitem{Varzielas:2012pa} 
  I.~de Medeiros Varzielas and L.~Lavoura,
  J.\ Phys.\ G {\bf 40}, 085002 (2013)
  doi:10.1088/0954-3899/40/8/085002
  [arXiv:1212.3247 [hep-ph]].


\bibitem{Grimus:2013rw} 
  W.~Grimus,
  J.\ Phys.\ G {\bf 40}, 075008 (2013)
  doi:10.1088/0954-3899/40/7/075008
  [arXiv:1301.0495 [hep-ph]].

\bibitem{Chankowski:1993tx} 
  P.~H.~Chankowski and Z.~Pluciennik,
  Phys.\ Lett.\ B {\bf 316}, 312 (1993)
  doi:10.1016/0370-2693(93)90330-K
  [hep-ph/9306333].


\bibitem{Babu:1993qv} 
  K.~S.~Babu, C.~N.~Leung and J.~T.~Pantaleone,
  Phys.\ Lett.\ B {\bf 319}, 191 (1993)
  doi:10.1016/0370-2693(93)90801-N
  [hep-ph/9309223].


\bibitem{Ellis:1999my} 
  J.~R.~Ellis and S.~Lola,
  Phys.\ Lett.\ B {\bf 458}, 310 (1999)
  doi:10.1016/S0370-2693(99)00545-6
  [hep-ph/9904279].


\bibitem{Fritzsch:1999ee} 
  H.~Fritzsch and Z.~z.~Xing,
  Prog.\ Part.\ Nucl.\ Phys.\  {\bf 45}, 1 (2000)
  doi:10.1016/S0146-6410(00)00102-2
  [hep-ph/9912358].


\bibitem{Xing:2000ea} 
  Z.~z.~Xing,
  Phys.\ Rev.\ D {\bf 63}, 057301 (2001)
  doi:10.1103/PhysRevD.63.057301
  [hep-ph/0011217].


\bibitem{Zhou:2014sya} 
  Y.~L.~Zhou,
  arXiv:1409.8600 [hep-ph].

\bibitem{Hagedorn:2010th} 
  C.~Hagedorn, S.~F.~King and C.~Luhn,
  JHEP {\bf 1006}, 048 (2010)
  doi:10.1007/JHEP06(2010)048
  [arXiv:1003.4249 [hep-ph]].


\bibitem{Hagedorn:2012ut} 
  C.~Hagedorn, S.~F.~King and C.~Luhn,
  Phys.\ Lett.\ B {\bf 717}, 207 (2012)
  doi:10.1016/j.physletb.2012.09.026
  [arXiv:1205.3114 [hep-ph]].


\bibitem{Ding:2013hpa} 
  G.~J.~Ding, S.~F.~King, C.~Luhn and A.~J.~Stuart,
  JHEP {\bf 1305}, 084 (2013)
  doi:10.1007/JHEP05(2013)084
  [arXiv:1303.6180 [hep-ph]].


\bibitem{Feruglio:2013hia} 
  F.~Feruglio, C.~Hagedorn and R.~Ziegler,
  Eur.\ Phys.\ J.\ C {\bf 74}, 2753 (2014)
  doi:10.1140/epjc/s10052-014-2753-2
  [arXiv:1303.7178 [hep-ph]].

\bibitem{Bazzocchi:2009pv} 
  F.~Bazzocchi, L.~Merlo and S.~Morisi,
  Nucl.\ Phys.\ B {\bf 816}, 204 (2009)
  doi:10.1016/j.nuclphysb.2009.03.005
  [arXiv:0901.2086 [hep-ph]].


\bibitem{Ding:2013eca} 
  G.~J.~Ding and Y.~L.~Zhou,
  Nucl.\ Phys.\ B {\bf 876}, 418 (2013)
  doi:10.1016/j.nuclphysb.2013.08.011
  [arXiv:1304.2645 [hep-ph]].


\bibitem{deMedeirosVarzielas:2017hen} 
  I.~de Medeiros Varzielas, T.~Neder and Y.~L.~Zhou,
  Phys.\ Rev.\ D {\bf 97}, no. 11, 115033 (2018)
  doi:10.1103/PhysRevD.97.115033
  [arXiv:1711.05716 [hep-ph]].

\bibitem{Morozumi:2017rrg}
  T.~Morozumi, H.~Okane, H.~Sakamoto, Y.~Shimizu, K.~Takagi and H.~Umeeda,
  Chin.\ Phys.\ C {\bf 42} (2018) no.2,  023102
  doi:10.1088/1674-1137/42/2/023102
  [arXiv:1707.04028 [hep-ph]].


\end{thebibliography}
\end{document}